A World unto Itself: Human Communication as Active Inference

Jared Vasil[1] (jared.vasil@duke.edu)

Paul B. Badcock[2,3,4] (pbadcock@unimelb.edu.au)

Axel Constant[5,6,7] (axel.constant.pruvost@gmail.com)

Karl Friston[7] (k.friston@ucl.ac.uk)

Maxwell J. D. Ramstead[6,7,8,9] (maxwell.d.ramstead@gmail.com)

AUTHOR AFFILIATIONS

1. Department of Psychology and Neuroscience, Duke University, Durham, NC, USA, 27708.

2. Centre for Youth Mental Health, The University of Melbourne, Melbourne, Australia, 3052.

3. Melbourne School of Psychological Sciences, The University of Melbourne, Melbourne, Australia, 3010.

4. Orygen, the National Centre of Excellence in Youth Mental Health, Melbourne, Australia, 3052.

5. Charles Perkins Centre, The University of Sydney, Camperdown NSW 2006, Australia.

6. Culture, Mind, and Brain Program, McGill University, Montreal, Canada.

7. Wellcome Centre for Human Neuroimaging, University College London, London, UK, WC1N3BG.

8. Department of Philosophy, McGill University, Montreal, Quebec, Canada

9. Division of Social and Transcultural Psychiatry, Department of Psychiatry, McGill University, Montreal, Quebec, Canada.

CORRESPONDING AUTHOR

Jared Vasil

Email: jared.vasil@duke.edu

Department of Psychology and Neuroscience

Duke University

Durham, NC 27708

WORD COUNT: 15,995 (main text, footnotes, figures); 14,226 (main text, footnotes); 13,037

(main text)

## Abstract (342 words)

Recent theoretical work in developmental psychology suggests that humans are predisposed to align their mental states with those of other individuals. One way this manifests is in *cooperative communication*; that is, intentional communication aimed at aligning individuals' mental states with respect to events in their shared environment. This idea has received strong empirical support. The purpose of this paper is to extend this account by proposing an integrative model of the biobehavioral dynamics of cooperative communication. Our formulation is based on *active inference*. Active inference suggests that action-perception cycles operate to minimize uncertainty and optimize an individual's internal model of the world. We propose that humans are characterized by an evolved *adaptive prior belief* that their mental states are aligned with, or similar to, those of conspecifics (i.e., that 'we are the same sort of creature, inhabiting the same sort of niche'). The use of cooperative communication emerges as the principal means to gather evidence for this belief, allowing for the development of a shared narrative that is used to disambiguate interactants' (hidden and inferred) mental states. Thus, by using cooperative communication, individuals effectively attune to a hermeneutic niche composed, in part, of others' mental states; and, reciprocally, attune the niche to their own ends via epistemic niche construction. This means that niche construction enables features of the niche to encode precise, reliable cues about the *deontic* or *shared value* of certain action policies (e.g., the utility of using communicative constructions to disambiguate mental states, given expectations about shared prior beliefs). In turn, the alignment of mental states (prior beliefs) enables the emergence of a novel, contextualizing scale of *cultural* dynamics that encompasses the actions and mental states of the ensemble of interactants and their shared environment. The dynamics of this contextualizing layer of cultural organization feedback, across scales, to constrain the variability of the prior expectations of the individuals who constitute it. Our theory additionally builds upon the active inference literature by introducing a new set of neurobiologically plausible computational hypotheses for cooperative communication. We conclude with directions for future research.

"*The point we emphasize is strong confidence in our original nature,*" Suzuki (1970/2014, p. 35)

## Introduction

An influential body of recent work on human communication describes it as *cooperative communication*. Cooperative communication is defined as intentional communication aimed at the alignment of mental states between conspecifics (reviewed in Tomasello, 2008, 2014, 2019). This is thought to be one particularly important behavioral manifestation of a broader, species-typical motivation to align mental states with those of others (Tomasello, Carpenter, Call, Behne, & Moll, 2005). Some have hypothesized that this motivation is the result of selective pressures acting on human evolution in the context of interdependent collaborative foraging (Tomasello, Melis, Tennie, Wyman, & Herrmann, 2012; Whiten & Erdal, 2012). In scenarios where individuals in a group must forage together for resources (food, water, information, etc.), the alignment of multiple individuals' goals, intentions, and attentional processes is necessary for success (e.g., Liebenberg, 2006). This view has been useful for empirical investigation in developmental and comparative psychology (reviewed in Call, 2009; Carpenter & Liebal, 2011; MacLean, 2016).

The purpose of this narrative review is to extend the approach to cooperative communication introduced above by leveraging a recent *active inference formulation* in theoretical neuroscience and biology (Friston, 2012, 2013). This formulation of living systems provides a formal account of the dynamics of belief-guided, embodied action from first principles of biological self-organization (e.g., Friston, Sengupta, & Auletta, 2014; Sengupta et al., 2016). A formal account is arguably important, because it forces one to make explicit one's theoretical predictions in experimental and modelling work that investigates the usage, development, and cultural evolution of human communication (e.g., Christiansen & Kirby, 2003; McCauley &

Christiansen, 2014). Furthermore, and although this is not the primary focus of this work, by proposing an active inference formulation of cooperative communication, we pave the way for a set of well specified predictions about the neurocomputational dynamics underwriting cooperative communication (Friston, 2010; e.g., Adams, Shipp, & Friston, 2013; Bastos et al., 2012; Parr & Friston, 2017, 2018). This is important, as precisely formulated neuroscientific hypotheses are largely absent from extant work on cooperative communication.

In brief, active inference is a mathematical formulation of the tendency of living systems to maintain themselves in a restricted set of states (i.e., their phenotypic states) while embedded in a fluctuating, partially observed environment (Friston, 2012, 2013). More precisely, active inference formalizes the structure of exchanges between organisms (individuals and groups) and their environment by explaining how the structure and function of organisms and their ecological niches become *attuned* to, or predictive of, each other (Bruineberg, Kiverstein, & Rietveld, 2018). In short, active inference suggests that every organism optimizes its internal (generative) model of the world via circular or self-fulfilling action-perception cycles that minimize an upper bound on biophysical surprise (i.e., variational free-energy). In turn, the environment becomes attuned to the organisms that inhabit it (Constant et al., 2019). We will see later that this is formally equivalent to maximizing the evidence for internal or generative models of the world – and that when the world (e.g., the cultural niche) is 'shared,' then the generative models of its denizens become committed to a (reliably) shared narrative.

Following a recent hypothesis of the embodied human brain derived from active inference, called the hierarchically mechanistic mind (Badcock, Friston, & Ramstead, 2019; Badcock, Friston, Ramstead, Ploeger, & Hohwy, 2019), our proposal combines active inference with substantive research in psychology and allied disciplines that captures the specific evolutionary, developmental, and real-time dynamics that underlie the human capacity for cooperative communication.

A key corollary of this approach is the construct of an *adaptive prior* (Badcock et al., 2019a,b). Adaptive priors are evolutionarily endowed, heritable beliefs[1] that guide characteristic patterns of cognition and behavior in conspecifics. In other words, adaptive priors have been shaped by selection to steer action-perception cycles toward adaptive, unsurprising outcomes (Badcock et al., 2019a; 2019b; Ramstead et al., 2018). Such priors depend upon genetic, epigenetic, and/or cultural inheritance, and often incorporate learned, *empirical priors* gleaned from experience to allow for sensitive adaptation to the local environment (Badcock et al., 2019b). Stated otherwise, adaptive priors effectively constrain the space of prior beliefs learned during ontogeny to enable adaptive action in local cultural niches (Badcock et al., 2019b; Ramstead, Constant, Veissière, & Friston, 2019).

Our proposal is as follows. We suggest that natural selection has endowed humans with an *adaptive prior for alignment*; i.e., an adaptive prior preference for action policies that generate sensory evidence that reliably indicates that their own mental states are aligned with, or similar to, those of conspecifics. This adaptive prior fosters intentional, patterned action sequences that gather *evidence* (i.e., sensory observations) for this belief; that is, that gather evidence for the hypothesis that 'we are the same kind of creature, inhabiting the same kind of niche.' The adaptive prior here functions to bias action and inference by leading agents to actively sample their sensorium in a way that, on average and over time, disambiguates conspecifics' (hidden) mental states. This sampling process is therefore *guided by*, and *generates evidence for*, the belief that our mental states are aligned. In short, we cast cooperative communication as an evidence gathering process; indeed, one that extends across temporally nested scales of analysis. The existence of this process follows from, and only from, an adaptive prior specifying the alignment of individuals' mental states[2]. Cooperative communication can thus be cast as a self-fulfilling prophecy, driven by the belief that we are alike. This belief is then characteristically reinforced by the evidence generated by belief-guided communication.

---

[1] Crucial to our account, in the active inference formulation 'beliefs' refer to (subpersonal) Bayesian beliefs – in the sense of Bayesian belief updating or belief propagation (as opposed to propositional beliefs).

Shweder and Sullivan wrote that "cultural psychology endeavors to understand how such divergences [in the processes that underwrite consciousness] relate to acts of interpretation and to the socially constructed meaning or representation of stimulus events" (1993, p. 506). The present article contributes to the project of cultural psychology and neuroscience (e.g., Han, 2015; and articles in the present collection) by explaining how a cultural milieu can shape and direct the dynamics of individual minds; and, in turn, how individual minds can shape their cultural milieu. We do this by providing an account of sociocultural cognition based on a shared adaptive prior for alignment, drawing on the active inference formulation. In turn, we argue that *how* one's cultural experience manifests in any given time and place – the particular tools one that uses in coming to grips with their world (i.e., words, gestures, and concepts) – is dependent on the history and current contingencies of one's culture and the minds, practices, and places that make it up.

The structure of the remainder of the paper is as follows. In order for readers to appreciate the broader context that underscores our proposal, we devote our second section to a review of some of the key phenomena that underwrite cooperative communication, as emphasized by other theorists to date. In the third section, we introduce relevant aspects of active inference, illustrated by examples drawn from studies of cooperative communication. In the fourth section, we leverage the background provided in the second and third sections to argue that human species-typical adaptive priors prescribe the alignment of one's mental states with those of conspecifics. This latter argument is presented in three subsections. The first subsection focuses on real-time dynamics (i.e., interaction) from the perspectives of an individual and dyad, respectively; the second focuses on ontogeny; and the third focuses on the timescale of cultural evolution. Our paper concludes with a few comments about the limitations of the current proposal of an adaptive prior for alignment. This is complemented with suggested directions for future research.

---

[2] On a deflationary view, this is the only solution that can exist, in terms of minimising the surprise or free energy of coupled free energy minimising agents. See below and Friston, Levin, Sengupta, & Pezzulo (2015) for a fuller discussion in the context of pattern formation.

## Theoretical Background

### The evolutionary origins of cooperative communication

Evolutionarily selected 'mutual expectations of cooperativeness' are thought to motivate the usage of cooperative communication (Tomasello, 2014). From the perspective of evolutionary biology, these expectations can be explained by considering the selective contexts that favored them. One promising candidate is so-called obligate collaborative foraging (Tomasello et al., 2012), where adaptive success in securing food and other resources is marked by a necessary dependence on cooperation with others (also, Baumard, André, & Sperber, 2013). For instance, in mutualistic 'stag hunt' games, a single individual is necessary to obtain a low risk, but low reward, food item (a hare), but two individuals are necessary to obtain a high risk, but high reward, food item (a stag). Here, collaboration appears as the riskier, but more rewarding, option[3]. It is riskier because, to cooperate effectively, the would-be partners must somehow align their mental states – their goals, intentions, and attention (Skyrms, 2001). Cooperative communication is thereby favored as a means to intentionally bring about the alignment of mental states. For instance, in high risk stag hunt scenarios preschool children communicated more, and more often, relative to low risk situations (Duguid, Wyman, Bullinger, Herfurth-Majstorovic, & Tomasello, 2014). Such joint foraging scenarios may point towards an important and recurrent aspect of the early selective pressures that favored the motivations and skills underlying cooperative communication (McLoone & Smead, 2014).

---

[3] In the active inference formulation, below, collaboration is 'rewarding' in the sense of maximising a shared or *prosocial utility* Devaine, M., Hollard, G., Daunizeau, J., 2014. Theory of Mind: Did Evolution Fool Us? PLOS ONE 9, e87619, Yoshida, W., Dolan, R.J., Friston, K.J., 2008. Game theory of mind. PLoS Comput Biol 4, e1000254..

Research examining the communicative behavior of extant non-human primates is crucial for understanding the evolutionarily nascent form of modern humans' communicative motivations and skills (Call & Tomasello, 2007; Mitani, 2009). Such work suggests that, generally speaking, the motivation and skills of non-human primates for intentional communication may have been gradually 'cooperativized' across human evolution (Tomasello, 2014); that is, exapted for both cooperative and competitive purposes with conspecifics. This trajectory may have begun with the usage of gestural communication geared towards simply eliciting specific responses from certain individuals (Call & Tomasello, 2007). For instance, something like ritualized great ape 'attention grabbers' – where an individual has learned that (for a certain conspecific) an action like slapping the ground loudly will likely bring about a desired state of the world (e.g., the initiation of play; Tomasello, 2008) – may have been the evolutionary precursor to certain manifestations of cooperative communication, like declarative pointing (Tomasello, 2019). Indeed, the motivational component is key (Rekers, Haun, & Tomasello, 2011): human-raised non-human great apes will occasionally point for humans (though never for conspecifics). However, they only do this 'selfishly,' that is, only when they expect the gesture to cause the individual to (say) get an out-of-reach object for the ape (Bullinger, Zimmermann, Kaminski, & Tomasello, 2011). In contrast, with cooperative communication, the underlying motive is argued to be 'fundamentally' cooperative (Tomasello, 2019); that is, from the onset of cooperative communication in ontogeny, human infants only appear satisfied following a communicative bid when their communicative partner has aligned their mental states with their own, with respect to the infant's intended referent (reviewed in Carpenter & Liebal, 2011; for comparative considerations, see Carpenter & Call, 2013).

## The developmental origins of cooperative communication

Human infants begin to use cooperative communication to align and coordinate mental states at nine to twelve months of age (Carpenter, Nagell, & Tomasello, 1998). This window of emergence in ontogeny is strongly maturationally constrained (Matthews, Behne, Lieven, & Tomasello, 2012), as evidenced by the emergence of communicative pointing at this age in every cultural setting studied (Callaghan et al., 2011; Lieven & Stoll, 2013; Liszkowski et al., 2012). One way this manifests initially is in declarative pointing gestures directed towards referents in the immediate environment. Experimental work suggests that the goal of infants' communication in such cases is to mutually align emotions, attitudes, and/or thoughts about a referent with another individual (Tomasello, Carpenter, & Liszkowski, 2007; e.g., Liszkowski, Carpenter, & Tomasello, 2007; Liszkowski, Schäfer, Carpenter, & Tomasello, 2009). Consistent with this, infants become disgruntled when others ignore their communicative bids for alignment. For instance, Liszkowski et al. (2004) found that infants became unsatisfied with uncooperative adults who ignored infants' communicative bids, who did not provide an emotional response symmetrical to the infant's, and who did not shift the focus of their attention back and forth between the infant and their referent.

This suggests that one aspect of the desired state of the world that motivates infants' earliest communication simply *is* alignment with other agents' mental states (Tomasello, Carpenter, & Liszkowski, 2007).

This example illustrates a signal feature of cooperative communication; namely, joint attention to a referent (Tomasello, 2008). There is substantial inconsistency in definitions of joint attention within and across psychological subdisciplines (Siposova & Carpenter, 2019). We follow the lead of Tomasello and colleagues (e.g., Tomasello, 1995) by defining *joint attention* as triadic situations in which two or more individuals possess reliable evidence that all participants are attending to the same referent, *and* that all participants know they are attending to the same referent (i.e., 'attending together'). This formulation of joint attention – in terms of reliable evidence for the mutually inferred alignment of attention (cf. mental states) – fits well with our proposal, which mandates the gathering of reliable evidence for the alignment of mental states.

The importance of joint attention for enabling cooperative communication comes from the fact that joint attention enables, and is enabled by, individuals' capacity to reliably 'ground' their communication in shared referents (Clark, 1996). Grounding creates something called common ground (Clark & Brennan, 1991). *Common ground* is the set of mental states (knowledge, beliefs, emotions, etc.) that is inferred to be reliably shared with others (Clark, 1996; Gadamer, 2003; Tomasello, 2014). The capacity to regulate communication with others by leveraging joint attention and common ground is present from the onset of cooperative communication (Tomasello et al., 2007). For instance, young infants use their shared experience with a particular person to comprehend and produce utterances and pointing gestures directed toward that individual (Ganea & Saylor, 2007; Liebal, Behne, Carpenter, & Tomasello, 2009; Liebal, Carpenter, & Tomasello, 2010; Saylor & Ganea, 2007; Tomasello & Haberl, 2003).

Moreover, part of regulating communication with respect to common ground is understanding, for instance, that one must try to 'fit' their communication to the inferred needs of another (Clark & Wilkes-Gibbs, 1986). As a simple example of this kind of 'perspectivizing' (Verhagen, 2007) or 'recipient design' (Schegloff, 2006) process, consider that how one chooses to talk about an artefact varies as a function of the inferred amount of cultural common ground shared with one's interlocutor. In the presence of much cultural common ground, a communicator might opt for brevity; and conversely, in the presence of less cultural common ground, one might use more precise (explicit, descriptive) language. For instance, when conversing with someone from Western cultural groups, one might employ the more cumbersome, longer descriptive utterance "the terrifying creature from Turkish folklore that appears during sleep paralysis" instead of the shorter proper name "Karabasan" (Jalal et al., in press). The upshot is that, in general, more common ground means less communication is needed to align mental states to a sufficient degree, and less common ground means more communication is required (Tomasello, 2008). In other words, the amount of information necessary to align mental states to a degree adequate to enable cooperative behavior within a given context is inversely proportional to the amount of common ground.

This turns on an important point: the optimization of relevance in cooperative communication (Sperber & Wilson, 1986). *Relevance* refers to the complexity-accuracy trade-off involved in the production and interpretation of communication; e.g., the trade-off between simplicity or compressibility, and meaningfulness or expressivity. A useful way to think about how this trade-off is finessed is in terms of communicative constructions (Goldberg, 2003). *Communicative constructions* are patterned pairings of form and meaning (e.g., word parts and order, intonation) whose synchronic use and form are the result of diachronic patterns of use and associated intergenerational transmission (e.g., processes of grammaticalization and reanalysis; Bybee, 2010; Heine & Kuteva, 2002). Cooperative communicators use communicative constructions to communicate (and thereby align their mental states).

Optimizing relevance, for a speaker, therefore means using the most minimal form that is expected to enable a listener to recover (something sufficiently similar to) the intended meaning (Kanwal, Smith, Culbertson, & Kirby, 2017); and for a listener, it means inferring the most parsimonious meaning that sufficiently explains the speaker's intentions (Kao, Wu, Bergen, & Goodman, 2014; see Goodman & Frank, 2016). This means, as above, that individuals sharing more common ground require less form to adequately align mental states, while those sharing less common ground require relatively more form (Winters, Kirby, & Smith, 2018). Relatedly, simpler propositions generally require less form to convey, and more complex propositions require more form (Kemmer, 2003). Producing and interpreting relevant communicative constructions thus has implications across the communicative signal, which spans from (e.g.) lexical selection and word order choice to the sequencing of particular phonemes and intonation patterns (Aylett & Turk, 2004).

Importantly, how might an individual recognize another's intention to generate an act of communication intended 'for' oneself in the first place (e.g., Behne, Carpenter, & Tomasello, 2005)? From another perspective, how might one make mutually apparent one's proximate motivation to align mental states, that one is communicating 'for' another individual? To this end, researchers have proposed that *ostensive cues* (Sperber & Wilson, 1986), like eye contact, spatiotemporal contingency, and the communicative (e.g., vocal) signal itself, play an important role in making mutually apparent an agent's intentions to communicate information intended to align mental states (reviewed in Csibra, 2010; indeed, Tomasello, 2014, synonymously calls cooperative communication 'ostensive-inferential' communication). Ostensive cues work by 'grabbing' the attention of others to redirect it triadically (i.e., towards the intended referent) so as to comment on it (Szufnarowska, Rohlfing, Fawcett, & Gredebäck, 2014). Thus, via their modulatory effects on the allocation of (joint) attention, ostensive cues play a critical (if indirect) role in increasing individuals' common ground and enhancing the reliability of one's inferences about this common ground (e.g., Moll, Carpenter, & Tomasello, 2007). This has important downstream effects on subsequent behavior. For example, communicative eye contact causes preschoolers to quickly infer another's desire to collaboratively play a stag hunt game (Siposova, Tomasello, & Carpenter, 2018; Wyman, Rakoczy, & Tomasello, 2013). Moreover, via their effects on attention, ostensive cues play an important role in guiding inductive inference and top-down categorization processes throughout ontogeny (Butler & Tomasello, 2016; Kovács, Téglás, Gergely, & Csibra, 2017).

**Taking stock**

In sum, five key components characterizing cooperative communication were noted in this section. Discussion of these components structures much of Section 4. First, great apes do not characteristically employ communication geared towards aligning mental states with conspecifics. Moreover, something like the motivations and skills underlying the communication of great apes likely served as a precursor to the evolution of cooperative communication in humans. Second, human communication is fueled by a motivation to align and coordinate mental states with conspecifics. This is a kind of mutual expectation of cooperativeness that is manifest most basically in processes of joint attention, which serves as a kind of 'evolutionarily endowed' common ground that gets the process of communication 'off the ground' in human ontogeny. Third, individuals using cooperative communication optimize the relevance of their communication, that is, the produced and inferred expressiveness of the communicative signal with respect to the production and processing costs of that signal. This depends on the common ground shared by interlocutors, such that, all else being equal, more common ground means less communication and less common ground demands more communication. Fourth, ostensive cues signal one's intention to communicate to another individual (and help one to disambiguate another's intention to communicative to oneself). These are cues like eye contact, contingency, and the speech signal itself.

Fifth and finally, it is useful to highlight that cooperative communication typically manifests, particularly in early ontogeny, as a circular or bidirectional flow of information (note, e.g., the double-arrowed base of the canonical 'joint attentional triangle'; Carpenter & Liebal, 2011). Thus, although we introduced cooperative communication by focusing largely on individual imperatives, it is a fundamentally collaborative process (Clark & Wilkes-Gibbs, 1986). The usage of cooperative communication is a relevance-optimized exchange of perspectives that manifests as a circular process of 'least collaborative effort' (Clark & Brennan, 1991). This characteristic circularity endows individuals with a single shared narrative constituted by their individual perspectives and roles in the collaborative. Figure 1 summarizes these points along with several others introduced in section 4.

Implicit in the preceding discussion is the idea that it would be *surprising* – that is, highly atypical – to find an adult human without a communicative system that they could employ to align their mental states with those of others. In this sense, the usage of cooperative communication is a predictable, or expected, aspect that characterizes part of the human phenotype. A question one might ask is, How does this expectation over species-typical states (i.e., this aspect of the phenotype) persist, robustly, across time and (action in) a fluctuating niche?

## Active Inference

Active inference is a theory of belief-guided adaptive action (Friston, FitzGerald, Rigoli, Schwartenbeck, & Pezzulo, 2017). It is a mathematical framework that models the processes by which organisms and their niche come to 'fit' or become 'attuned' to each other (for an introduction to the mathematical apparatus of active inference, see Bogacz, 2017; Buckley, Kim, McGregor, & Seth, 2017). In other words, active inference describes the manner in which organisms and their environments come to possess statistical properties that are predictable from each other (Bruineberg, Rietveld, Parr, van Maanen, & Friston, 2018; Constant, Ramstead, Veissière, Campbell, & Friston, 2018). On this view, organisms come to embody statistical models of their ecological niche via perception and learning, and both cultural and natural selection (i.e., empirical and adaptive priors, respectively). Reciprocally, organisms modify their niche to fit their prior beliefs via adaptive action and niche construction. As detailed in Figure 2, the models in this formulation are 'generative models' that recapitulate the causal independencies between the factors that generate their sensory input (i.e., how the niche causes their sensory data; e.g., Hinton, 2007; LeCun, Bengio, & Hinton, 2015). In active inference, organisms are, roughly speaking, normative models of what *ought* to be the case, given 'the kind of creature that I am' (Friston, 2011).

The main theoretical suggestion of this paper is that human individuals appear, characteristically (i.e., species-typically), to be endowed with an *adaptive prior that one's mental states are aligned with those of conspecifics*. Now, for human agents, the mental states of other agents are unobservable or 'hidden' states that need to be inferred on the basis of perceptual cues (e.g., gaze direction, posture, facial expression). In other words, mental state alignment is an inference problem: to align with others, an agent must infer the latent or hidden causes (i.e., mental states) that generate observable consequences (i.e., actions). Thus, for agents whose niche includes the mental states of other agents, the set of actions that resolve uncertainty about the niche must comprise actions that reliably disambiguate others' mental states[4]. We suggest that this is precisely the situation brought about by the presence of an adaptive prior for alignment. This adaptive prior fosters specific, patterned forms of (communicative) action and inference that are aimed at disambiguating the mental states of other agents. The characteristic result of this process is the alignment of mental states between conspecifics. The alignment process enables and maintains reliable hypotheses about shared narratives that contextualize our experience (Friston & Frith, 2015a).

## Active inference, adaptive priors, and alignment

In active inference, actions are generated by hierarchically organized *policies* (beliefs about action). The policy pursued by an organism at a particular time is the one that minimizes an information-theoretic *variational free energy* term (Friston et al., 2015; for a review of variational inference, see Blei, Kucukelbir, & McAuliffe, 2017). Roughly speaking, free energy quantifies the discrepancy between what an agent expects or prefers to sense and what it actually senses. This conception of free energy is closely related to prediction error (i.e., the mismatch between predicted and observed sensations; Clark, 2013). A complementary view of free energy is that it scores the (negative log) evidence for the internal model generating predictions, in the sense that sensory data that conform to predictions provide evidence for the veracity of the agent's generative model. In short, minimizing free energy is the same as soliciting sensory evidence for one's model of the world (sometimes known as 'self-evidencing'; Hohwy, 2016). On this view, we are our own existence proofs.

---

[4] This is described anthropomorphically, as though individuals are explicitly engaged in an inference like 'I want to attune the statistics of our brains.' Clearly, this is not the case. Rather, this language is used for expository purposes. Indeed, this descriptive tendency is essentially the same as that used by, e.g., Tomasello (2019), where talk of humans' ultimate motivation to align and coordinate mental states often surfaces in place of proximate instantiations thereof. We thank L. Li (personal communication, December, 2018) for noting this ambiguity.

The free energy expected under a policy tracks the probability of that particular policy being pursued (i.e., of that specific policy being selected to guide action). Relatively less expected free energy indicates a relatively more probable policy (Friston et al., 2015; Pezzulo, Rigoli, & Friston, 2018; relatedly, Cisek, 2007). Expected free energy can be decomposed into two terms: *epistemic value* (the information gain of an observation), and *pragmatic value* (the expected log evidence of some outcome, given a generative model of how outcomes depend on action). The relative influence of each term quantifies the degree to which a particular policy generates actions that explore the niche (i.e., exploration), or actions that leverage reliable expectations about the niche to secure preferred outcomes[5] (i.e., exploitation) (Friston et al., 2015). This is depicted in Figure 2[6].

*Salient* policies are those that have a high epistemic value or affordance (Parr & Friston,

---

[5] These outcomes are *a priori* preferred because they are the least surprising ones; i.e., outcomes that 'a creature like me' would expect to encounter.

[6] A mathematically equivalent but complementary division of expected free energy is into *ambiguity* and *risk*. Ambiguity is the expected uncertainty about outcomes given some state in the future, while risk is the divergence between anticipated and preferred outcomes. Interestingly, risk is also the expected *complexity cost* found in statistics. This means that minimising expected free energy – by selecting the right kind of policies – implicitly minimises the complexity cost of inference. This is exactly the imperative established in the previous section; namely to find 'common ground' that minimises communication cost. This particular perspective can be traced back to the foundational principles of universal computation, where variational free energy is often discussed in terms of minimum description or message lengths. See MacKay (1995), Schmidhuber (2010), and Wallace & Dowe (1999) for more discussion.

2017). These energize actions that enable an agent to learn the statistical regularities of its environment (see caption, Fig. 2). This, in turn, enables pragmatic imperatives to foster actions that capitalize on learned regularities (Friston et al., 2015). For example, repeated exposure to the sensory phenomena characteristic of their culture's communicative constructions leads infants to become familiar with the statistical properties of those constructions (Romberg & Saffran, 2010). In turn, increasingly precise expectations about hidden causes may lead infants to prefer gathering information from speakers of their native language relative to speakers of a foreign language (e.g., Begus, Gliga, & Southgate, 2016; Marno et al., 2016). This is predicted by the hypothesis that agents exploit their familiarity with the sensory phenomena characteristic of their culture's communicative constructions to guide attention towards sensory stimuli that is expected to be useful for disambiguating the mental states of others (Figures 2 and 3).

In active inference, the folk-psychological term 'attention' refers to two distinct, but closely related, phenomena; namely, *epistemic value* and *precision weighting* (Parr & Friston, 2017). Epistemic value, salience, or affordance is the component of policy selection just discussed; it is that component of the value of policies that tracks how much a policy reduces uncertainty about the state of the world (e.g., Friston, Adams, Perrinet, & Breakspear, 2012). It provides a description of the folk-psychological phenomenon of activity orienting towards or 'turning one's attention' to a certain modality or part of the sensory field (e.g., in visual saccades that sample a particular location in visual space). In short, salience or epistemic affordance is an attribute of how we sample the world – in the sense that actively sampling sensory information will reduce uncertainty, in relation to our current beliefs. In contrast, precision is an attribute of the sensory data *per se*. Imprecise sensory data should have less effect on (Bayesian) belief updating, relative to precise information. It is therefore important to afford the right precision to each sensory sample, via precision weighting.

Precision-weighing is the related (but distinct) attentional process that determines the relative influence of bottom-up error signals and top-down expectations in the brain; e.g., a high precision on sensory signals corresponds to low confidence in top-down beliefs (Clark, 2013; Powers, Kelley, & Corlett, 2016). That is, in the sense of precision-weighting, 'attention' refers to the optimization of the precision (inverse variance) of prior beliefs about the causes of sensory data, relative to the precision of those data; in other words, attentional selection is in the game of selecting the right sort of sensory information for belief updating. This precision weighting in the brain is thought to be mediated by the modulation of neuronal gain (Kanai, Komura, Shipp, & Friston, 2015). Precise (attended, ascending) error signals then serve to modulate action and direct what is learned (Adams et al., 2013; Feldman & Friston, 2010). The complement of this attentional selection is the attenuation of precision; known in psychophysics as *sensory attenuation*; i.e., attending away from or ignoring certain sensations; particularly those we cause ourselves.

Crucially, selective attention and attenuation of precision can be part of the covert (mental) actions that are entailed by a policy. In other words, when selecting the policy that minimizes expected free energy we are also committing to both overt action on the (embodied) world – through moving, blushing, speaking etc. – *and* a covert attentional set. We will now illustrate these aspects (orienting to salient stimuli and attentional selection) of active inference with two examples.

As a first example, in the case of human communication, orienting to salient sensory streams should enhance the ability to learn the causes (i.e., mental states) generating sensory evidence by making beliefs about mental states generating that stream more probable. With this in mind, note that one common motivation for infants' and young children's communication is quintessentially uncertainty resolving and 'interrogative' (Begus & Southgate, 2012; Harris, Bartz, & Rowe, 2017). For instance, infants' pointing can function as a request for information about the name or function of objects (Begus & Southgate, 2012; Kovács, Tauzin, Téglás, Gergely, & Csibra, 2014). It is thus interesting that, in line with the present account, orienting to communicative bids enhances learning of (e.g.) communicative constructions and object functions (Begus, Gliga, & Southgate, 2014; Lucca & Wilbourn, 2018a, 2018b; see Friston, & Frith, 2015a). In short, infants evince sophisticated policies for resolving uncertainty and creating opportunities for epistemic foraging. In turn, attending to and learning the causes of the communicative stream then enables policies to exploit prior beliefs about how such sensations were caused; that is, inferring whether or not we are aligned, based on the evidence generated through our interactions (e.g., in using learned constructions to ask, explicitly, 'Do you understand?'). This brings us to our second example.

For agents who expect their predictions to be fulfilled, individuals who do not provide evidence for this expectation – despite one's attempts to actively attune mental states – should come to be treated as imprecise sources of sensory information, relative to others that fulfil their expected 'role' in the evidence gathering process; i.e., others that are afforded epistemic trust (Fonagy and Allison, 2014). In other words, in a given communicative interaction, salient policies are those that are expected to be useful with respect to the alignment of mental states; e.g., in certain instances of conversational repair (Schegloff, Jefferson, & Sacks, 1977). Across interactions with specific others, repeatedly experiencing surprising responses (i.e., insufficient evidence for, or evidence against, alignment) means that selective attention towards those specific others comes to be afforded low precision (i.e., ignored). Subsequently, action should lead the appearance, on average across time, of avoidance such unreliable parts of the niche (Constant et al., 2018) – much as we tend to avoid the dark when searching for something (Demirdjian et al., 2005).

We suggest that this provides an explanation of the findings by Liszkowski et al. (2004), discussed above, which reported that 12-month-olds were dissatisfied with an uncooperative adult who failed to provide both look-backs between the infant and their intended referent and the same emotional response as the infant in response to the infant's communicative bids. On our view, infants were attempting a kind of fast 'error correction' by generating actions expected to minimize exposure to unexpected cues (i.e., allostatic control; see Pezzulo, Rigoli, & Friston, 2015). This occurred via a rapid increase in the salience of policies that generate pointing behavior when sampling sensory data that was inconsistent with infants' prior beliefs about alignment. Moreover, only the group of infants who attempted to communicate with an uncooperative adult pointed significantly less across trials; through the lens of active inference, they had revised their expectations about the sensory effects of action, leading them to select other policies.

This second example suggests that, within and across trials of the experiment, infants appeared to climb an evidence gradient for their expectations. That is, repeated orienting to cues indicative of the (dis)alignment of prior beliefs – despite allostatic control geared towards avoiding such surprising encounters – caused infants to infer and learn that their interaction partner was unhelpful with regards to gathering evidence for their (species-typical) prior beliefs. For the infant, orienting to the sensory consequences of repeated failed attempts to elicit evidence from the adult indicative of alignment (e.g., look-backs and symmetrical emotions) had an impact on the expected free energy of policies. In particular, policies geared towards inferring the prior beliefs of the uncooperative adult came to be characterized by a relatively high expected free energy.

Consequently, such policies became relatively unlikely to gain control over action; i.e., less communication with that adult.

In sum, by suggesting that humans are characterized by an adaptive prior for alignment, we effectively argue that policies expected to disambiguate others' mental states are characterized by a low expected free energy. This is by virtue of their high epistemic affordance (i.e., in a niche partly constituted by others' mental states). Consequently, these policies tend to dominate action – people tend to gesticulate and talk with others. Repeatedly leveraging this belief to guide context-sensitive patterns of action, in turn, enables agents to learn the structure and dynamics of their niche. Because the human niche includes others' mental states, beliefs about how to act to effectively infer and align with others will have high adaptive value (Constant, Clark, Kirchhoff & Friston, in press). This means that learning likely entails refining one's set of 'communicative policies' to approximate the set of policies expected (i.e., typically used) in one's cultural milieu. In short, leveraging communicative constructions means converging on the mutually inferred, or deontic, value of policies geared towards disambiguating mental states among agents equipped with an adaptive prior for alignment.

**Deontic value: Shared expectations about the value of policies**

Above we assumed that the prior beliefs of conspecifics had converged on the set of constructions leveraged in their cultural niche. This assumption is important, as our argument considers the acquisition and (cultural) evolution of communicative constructions (Sections 4.2 and 4.3). Within active inference, the concept of *shared* or *deontic value* – and associated *deontic cues* – (Constant et al., 2018, 2019) may be useful for understanding the emergence of cooperative communication in ontogeny and cultural evolution.

The deontic value of a policy rests on a direct ('automatized') likelihood mapping between learned cues and associated action policies. The mapping from deontic cue to policy is 'direct' in the sense that observation of a deontic cue comes to 'automatically' elicit an associated (i.e., learned) policy[7]. Deontic cues are observations that trigger such automatic, or habitual, policy selection (Constant et al., 2019). Encultured agents learn deontic observation-policy mappings in development, through their engagement with the deontic cues that populate their local cultural niche (e.g., Chukoskie, Snider, Mozer, Krauzlis, & Sejnowski, 2013). By 'offloading' cognition into the environment in this way (see Clark, 2008; and Clark, 2006), the direct mapping enables individuals to bypass costly updates to, and metabolic upkeep of, their beliefs about what to do (given what is inferred of the niche). This allows agents to rely directly on deontic cues to select the most appropriate policy (Constant et al., 2018). There is clearly a close relationship between deontic cues, semiotics, and signs (Goodwin, 2000; Sewell, 1992; Silverstein, 2003) that underwrite communication. Perhaps the most celebrated system of encultured deontic cues is language itself.

For instance, consider an individual who has learned the English construction 'let alone' (Fillmore, Kay, & O'Connor, 1988); that is, a communicative construction marked by a comparative 'let alone' phrase centered between clause X and clause (fragment) Y; e.g., 'I could barely run 1 mile let alone 4 miles'. Learning the 'let alone' construction, as one example of a more general phenomenon (Section 4), entails learning the deontic value of cues (for policies that parse spoken or written language). In short, if I hear you utter the phrase 'X' and possess prior, reliably shared knowledge of the construction 'X let alone Y', then I can reliably expect you to follow up with 'let alone Y.' This example assumes a probabilistic (generative) model of how communicative sensations are caused (e.g., a scheme to reliably parse syntax; Levy, 2008; reviewed in Kuperberg & Jaeger, 2016). In particular, this turns on the acquisition of the deontic value of linguistic policies entailed by the hypothesis that one is witnessing a 'let alone' construction.

---

[7] Technically, expected free energy is combined with deontic value to score the likelihood of a particular policy. In the absence of deontic value, the expected free energy will select the most apt epistemic and *goal directed* policy, given beliefs about the current state of the world. Conversely, if certain cues render the deontic value of a policy sufficiently high, it will dominate policy selection – and emerge as a *habit*.

But how do such reliable mappings come to exist in the first place? That is, how do communicative constructions 'build up' over (neurodevelopmental or evolutionary) time? Consider a simple example: continually walking along the same path across a park each day wears down the grass along that path (Constant et al., 2018). As the grass wears down and a clear path forms, one learns to expect the associated sensory cues when revisiting the path. Because of this, the path becomes increasingly salient for both oneself and for others 'like me', who can (like me) leverage such 'meaningful' traces left by my actions at later time points. Consequently, the cognitive processing associated with answering the question 'Where ought I to walk next' is afforded directly by physical features of the niche. This saves on the costs associated with planning as active inference (Attias, 2003; Baker and Tenenbaum, 2014; Botvinick and Toussaint, 2012; Mirza et al., 2016) – the inference is literally 'offloaded' into the environment (see equations in Figs. 2 and 3). The niche provides a clue as to what to do, reliably, as a deontic cue.

Crucially, when this process of 'carving out' deontic *cues* in the niche is performed by an increasing number of agents, the deontic value of policies and associated cues becomes increasingly robust to perturbations. In other words, the expectations of the social niche – here, the set of form-meaning pairings constituting a communicative system – become increasingly precise with increases in the number of interactions between agents constituting that system (Constant et al., 2019). Increasingly precise, niche-based expectations mean that agents become more likely to sensitize their behavior to that cue; e.g., the dynamics of a cue become sufficiently precise so as to enable learning of that cue and its associated action policy in ontogeny (see Section 4.3). In multi-agent systems equipped with an adaptive prior to align mental states, learning of deontic value (i.e., inferring the most common policies undertaken by other denizens of the niche) is learning the 'shared' value of a policy – the value of a policy for people 'like me' in our community (Figure 3).

What might it mean to offload cognition into the environment in the fashion above, for agents equipped with an adaptive prior to attune mental states? Individuals effectively outsource solutions to the problem of 'How ought I to talk' to the niche itself. The traces left by repeatedly aligning mental states via communication may enable the niche to subsequently afford increasingly precise, shared expectations about how other agents 'like me' (should) act so as to align mental states most effectively (e.g., during evolutionarily relevant stag hunt scenarios; Grau-Moya, Hez, Pezzulo, & Braun, 2013). In principle, this takes pressure off inferring "what sort of person am I in this context" (Moutoussis et al., 2014). Technically, it finesses the computational cost of belief updating from (deontically installed) priors to posterior beliefs about behaviors that are apt for the current setting. Consequently, the cue (or sequence of cues) may come to be preferred by both individuals during subsequent interactions in similar contexts (Lewis, 1969; Schelling, 1960; e.g., Clark & Wilkes-Gibbs, 1986). Formally speaking, at later instances of interaction, the expected free energy of historically selected policies – leveraged to align mental states – falls; such policies then tend to be selected to generate predictable action sequences geared towards the alignment of mental states (Friston & Frith, 2015b).

### Human Communication as Active Inference

This section provides a discussion of our proposal. The species-typical motivation to align mental states with conspecifics is cast as an adaptive prior preference for alignment. This, we suggest, provides the basis for a normative framework for predicting, explaining, and modelling the behavioral, psychological, and neural underpinnings of cooperative communication. Our discussion in this section telescopes from considerations at the microscale (i.e., interaction), to the mesoscale (i.e., ontogeny), and, finally, to the macroscale (i.e., cultural evolution).

### Dynamics at the timescale of interaction: The individual in context

A central part of the content of the prior for alignment is that the actions of agents (e.g., oneself) update the mental states (prior beliefs) of other agents. Because mental states cause action (and, hence, observations), gathering reliable evidence for this prior means that agents orient to the individual(s) towards whom their action is directed – the sensory consequences of one's action are realized by the actions of others. That is, if one expects to infer others' mental states, the only evidence available is found in the observed consequences of others' actions (Figure 4). Indeed, policies that direct action towards others – so as to disambiguate their mental states (e.g., attentional orienting and, later, cooperative pointing) – possess an evolutionarily unique (Seyfarth & Cheney, 2003), maturationally constrained salience from early in life (Matthews, Behne, Lieven, & Tomasello, 2012; Reddy, 2003). Gathering evidence for these expectations manifests in *coupled action-perception cycles* (Friston, & Frith, 2015a); i.e., intentionally co-constructed loops of action-perception that induce a reliable statistical coupling between two coupled agents (reviewed in, e.g., Feldman, 2015; Hasson & Frith, 2016; Hasson, Ghazanfar, Galantucci, Garrod, & Keysers, 2012). For expository purposes, we may say that, within the coupled action-perception cycle of human agents, evidence for the self amounts to evidence for the other; and evidence for the other is evidence for the self.

Above, we discussed how mutual expectations of cooperativeness play a crucial role in getting cooperative communication off the ground in ontogeny. This just means that the (epigenetically and neurodevelopmentally) constrained, precise beliefs about the similarity of others and oneself enable nascent individuals to engage in cooperative communication. In particular, such couplings are only possible because both agents possess reliable expectations that the other agent is sufficiently 'like me' (cf. Meltzoff, 2007): we share the same prior beliefs to attune hidden dynamics. This provides an initial 'naive' confidence in beliefs about how one's action will influence another's prior beliefs (that, in turn, influence sensory outcomes via their actions). Borrowing from the language of social constructivist views of development (e.g., Rhodes & Wellman, 2017), our prior is a kind of naive certainty in one's intuitive theory about agential efficacy, with respect to the mental states of others (see also Kelso, 2016). This is to say that prior beliefs about the niche, e.g., others' mental states, bottom out just in their expected free energy. Belief-guided action (e.g., collaboration) may thus be constrained by salient policies entailed by a prior belief that, psychologically speaking, some hypothesis is in common ground. Put simply, to the extent that this hypothesis is sufficiently reliable, it will guide action and inference (see Figure 4 and, e.g., Gallagher & Allen, 2018; Yoshida, Dolan, & Friston, 2008).

Pursuing this line of reasoning further provides a single, formally specified framework to subsume distinct proximate motivations for communication. That is, proximate motivations for communication (e.g., declarative, expressive, informative, interrogative motives; Begus & Southgate, 2012; Tomasello, 2019) surface as particular psychological manifestations of the same, species-typical tendency to align prior beliefs. Consider two proximate motivations for communication noted above; namely, a 'declarative' one motivated by the desired alignment of attentional states; and an 'interrogative' one motivated by a desire to learn about the niche. In the former case, individuals exploit their reliably shared beliefs to render the niche sufficiently similar to themselves (e.g., 'By ostensively pointing for that other agent, I expect to effectively align our mental states with respect to my intended referent'); and in the latter, individuals explore the precise, reliable parts of the niche (here, other agents) to improve their internal model of the niche (e.g., 'What is this thing called?'; reviewed in Harris, 2011; Harris et al., 2017). The underlying commonality in both cases is that individuals are effectively generating action-perception cycles that couple them to others, with the result being the alignment of mental states with respect to the niche.

Moving now to relevance optimisation, we remind the reader that this process involves finessing the trade-off between the accuracy (e.g., meaningfulness, expressivity) and complexity (e.g., minimum description length, hierarchical depth of the policy) of their communicative constructions. Under active inference (see Pezzulo, Donnarumma, & Dindo, 2013), if the prior beliefs of two individuals are inferred to be highly divergent on the basis of the evidence each provides to the other, and if both expect to minimize this divergence to a sufficient degree, then costlier (e.g., hierarchically deeper or more complex) policies should become relatively more salient as agents become increasingly dissimilar, as these policies will be necessary to resolve uncertainty or disambiguate the mental state of inscrutable others. This is in contrast to two individuals who 'speak the same language'. Here, less information needs to flow within the coupled action-perception cycle to attune mental states to a similar degree. In support of this view, one study (Kanwal et al., 2017) found that adults optimize the relevance of their communicative constructions during collaborative tasks as a function of their common ground, by using shorter words for common objects and longer words for uncommon objects (see also Winters et al., 2018). Related work suggests that children's adjective use (Bannard, Rosner, & Matthews, 2017), turn-taking dynamics (Butko & Movellan, 2010), and question asking (Nelson, Divjak, Gudmundsdottir, Martignon, & Meder, 2014) may be usefully cast as if they were optimizing the information content of produced communicative constructions with respect to processing and energy concerns (cf. Pea, 1979).

For a receiver, attention to the communicative stream enables updates to one's beliefs by providing 'contextual effects' (Sperber & Wilson, 1987); that is, orienting to a speaker influences the precision of hypotheses (about, e.g., the interpretation of an utterance) through appropriate selection of ascending sensory information (indexed neurophysiologically by alpha suppression; Höhl, Michel, Reid, Parise, & Striano, 2014; and increased theta; Begus et al., 2016; Köster, Langeloh, Höhl, 2019). Specifically, individuals appear to explain away incoming sensory data by zeroing in on informative (useful) but parsimonious (i.e., efficient) explanations of hidden causes (Goodman & Frank, 2016; see also Gershman, Horvitz, & Tenenbaum, 2015). For instance, Frank & Goodman (2014) report that adult and child listeners disambiguate ambiguous word meanings by optimizing their inferences of the relevance of a speaker's intended meaning[8]. In particular, these inferences can be captured as if individuals were maximizing model evidence for the prior belief that speakers are informative (see also Kao et al., 2014). This is captured by our extended formulation of cooperative communication, where inferences about mental states can be cast in terms of maximizing Bayesian model evidence (i.e., minimizing variational free energy) for the causes of one's sensation (e.g., another's mental states; Friston & Frith, 2015a).

---

[8] Interestingly, in machine learning, *automatic relevance determination* is a term used to denote model selection based upon variational free energy; namely, the removal of redundant model parameters to maximise efficiency or Bayesian model evidence. In turn, this is closely related to principles of minimum redundancy and maximum efficiency in perception (Barlow, 1974; Linsker, 1990; Wipf & Rao, 2007).

Given an adaptive prior for alignment, one should tend to favor policies expected to reliably generate evidence of engagement in a coupled action-perception cycle. That is, such *ostensive policies* – policies expected to generate ostensive cues – are adaptive because they tend to generate sensory evidence for the hypothesis that one is engaged in a coupled action-perception cycle. Ostensive policies indicate to one's communicative partner that attending to one's action (i.e., to the individual generating ostensive cues) will likely be informative for them. Consequently, for a recipient, evidence provided by such cues increases the salience of certain policies; e.g., attentional orienting geared towards disambiguating the speaker's prior beliefs (Szufnarowska et al., 2014). As attention optimizes the precision of sensory cues, ostension in the coupled action-perception cycle plays a crucial (if indirect) role in reliably entraining and shaping prior beliefs (Axelsson, Churchley, & Horst, 2012; Butler & Tomasello, 2016; Kovács et al., 2017). Since prior beliefs generate action, ostensive cues are thus critical for guiding other individuals' actions and hence one's (attended) sensory states (e.g., Siposova et al., 2018).

By the same logic, in response to ostensive cues a recipient should (ostensively) signal their own inferred entrance into a communicative coupling (e.g., uptake signals; Austin, 1962); as well as, for example, their subjective degree of (and certainty in) the attunement of mental states (e.g., backchannel signals; Clark & Brennan, 1991). Indeed, other individuals – inferred to possess the same adaptive prior for alignment – preferentially leverage cooperative communication in turn; that is, respond to one's communicative bids (Kishimoto, Shizawa, Yasuda, Hinobayashi, &

Minami, 2007; Wu & Gros-Louis, 2015). This makes sense in light of the adaptive prior specified here: responding to another's communicative bids is something in the interest of both agents[9].

In summary, this subsection provided an active inference account of the microscale features of cooperative communication, from an individual's perspective, noted in Section 2. We have thus outlined some important means by which individuals intentionally align their prior beliefs with respect to the dynamics of the niche (Constant et al., 2018), including others' mental states (Friston & Frith, 2015a). Indeed, a foundational facet of our account is that the alignment of the mental states of conspecifics manifests in the emergence of a novel scale of social and cultural dynamics constituted by synchronized component individuals (Ramstead et al., 2018). We turn to this now.

## Dynamics at the timescale of interaction: The dyad

The precision of one's prior beliefs relative to another agent's, with whom one is coupled, has important implications for the degree and direction of attunement within and across couplings. In particular, the relative precision of the prior beliefs of each agent constrains the characteristic pattern of information flow between them – both at the level of turn taking in dialogical exchanges, and at the level of learning useful generative models of others[10] (and implicitly, of the self) (Friston & Frith, 2015a; 2015b; Gencaga, Knuth, & Rossow, 2015; e.g., Schippers, Roebroeck, Renken, Nanetti, & Keysers, 2010). In terms of learning, this means that individuals endowed with relatively imprecise prior beliefs tend more, on average across time, to modify their own structure to fit that of their communicative partner(s), relative to individuals with relatively precise priors. This is a special case of generalized synchronization that is underwritten by the enslaving principle from cybernetics (Tschacher & Haken, 2007). To attune prior beliefs in such 'asymmetric' couplings, individuals with imprecise expectations in effect increase the precision of their sensory states (i.e., 'up the gain' afforded to sensory input; Auksztulewicz et al., 2017; Moran et al., 2013). This allows them to better change their own prior beliefs as a function of the evidence generated by their own (and others') action. This captures, for instance, the characteristic flow of information between agents following exposure to cues of prestige, with prestigious individuals being 'trend-setters' and others following suit (Henrich & Gil-White, 2001; Veissiére et al., 2019).

---

[9] It is useful to note that ostensive policies are salient insofar as they are (but one) intentional means for rapidly increasing the precision of (certain kinds) of hypotheses for another agent; e.g., that it is likely worthwhile to attend to the individual generating the ostensive cues. The account on offer therefore accommodates evidence suggesting that non-ostensive (unintentional) but nonetheless attention-grabbing actions, like shivering, may have similar effects on others' attentional orienting as ostensive cues (de Bordes, Cox, Hasselman, & Cillessen, 2013; Szufnarowska et al., 2014).

[10] In numerical analyses of coupled communication, *turn taking* is usually implemented by a reciprocal augmentation and attenuation of sensory precision – so that one member of the dyad is listening while the other is speaking. Please see Friston & Frith (2015a) more details. A more enduring asymmetry relates to how one can learn from others, as illustrated using simulations of birdsong in Friston & Frith (2015b).

Additionally, such an asymmetry in information flow may capture the dynamics of the coupled action-perception cycles characteristic of interactions between human infants and children, and adults. Experimental and computational evidence suggests that older individuals possess relatively precise prior expectations, relative to those of younger, less experienced individuals (Karmali, Whitman, & Lewis, 2018; Wolpe et al., 2016). Thus, younger individuals may ascribe greater precision to sensory information (Moran, Symmonds, Dolan, & Friston, 2014). The hypothesis here is, then, that repeated couplings between infants and children with adults (and more experienced peers) may cause the prior beliefs of inexperienced individuals to converge more towards the hidden causes generating sensory consequences (i.e., the mental states of more experienced others), rather than the other way around (Friston & Frith, 2015b; e.g., Fotopoulou & Tsakiris, 2017). That is, coupled action-perception cycles in such dyads tend to be characterized by an *asymmetric entrainment of prior beliefs* (for a closely related view, see Brownell, 2011).

What does this mean for the dynamics of (neural) belief updating during interaction? Technically, attunement to the niche instantiates the generalized synchronization of the statistics of prior beliefs and the niche (e.g., others' mental states); such that the structure and dynamics of individual brains come to recapitulate the structure and dynamics of the niche in which they are embedded[11] (Friston, 2012). This is depicted in Figure 5. Synchronization is a phenomenon that occurs in coupled chaotic dynamical systems (Pecora, Carroll, Johnson, Mar, & Heagy, 1997). Technically, it means that there is a (diffeomorphic) function relating the dynamics of the state of one system to those of the system with which it is coupled (Pecora, Carroll, & Heagy, 1995). For instance, modelling results suggest that endowing two coupled hierarchical dynamical systems with an expectation to infer the hidden causes generating another's actions enables a bidirectional flow of information that synchronizes the statistics of their prior beliefs (Constant et al., 2018; Friston & Frith, 2015a). Alignment within and across coupled action-perception cycles means that the similarity (technically, the mutual information) of individuals' expectations increases (Friston & Frith, 2015b; Hasson & Frith, 2016). In this scheme, attention functions as a kind of coupling parameter, and its allocation is constrained by adaptive priors. Attention effectively increases the amount of information transferred from the system with precise priors to the system with imprecise priors (i.e., the system increasing the gain of its sensory states).

Indeed, studies in 'two-person' or hyperscanning neuroscience (Schilbach et al., 2013) have found evidence of the synchronizing effects of the usage of cooperative communication during, e.g., unidirectional person-to-person monologues (Liu et al., 2017; Pérez et al., 2017;

---

[11] Technically, hidden states are characterized by their sufficient statistics. This denotes the minimum quantities needed to fully describe a probability distribution (Cover & Thomas, 1991). For the Gaussian distributions used in active inference, these are the mean and variance (see Buckley, Kim, McGregor, & Seth, 2017). Generalised synchronisation implies that the mutual information of (the dynamics of) the states occupied by two (e.g.) chaotic dynamical systems is high (Pecora et al., 1995).

Stephens, Silbert, & Hasson, 2010), person-to-group monologues (Schmälzle, Häcker, Honey, & Hasson, 2015), bidirectional person-to-person dialogues (Jiang et al., 2012), and even between classmates and their teacher during daily school activities (Dikker et al., 2017). Crucially, the degree of interbrain synchrony of neural dynamics appears to strongly predict psychological phenomena; for instance, the subjective meaningfulness of communication (Stolk et al., 2014), the accuracy of recall of the content of communication (Zadbood, Chen, Leong, Norman, & Hasson, 2017), and the perceived 'power' of political speech (Schmälzle et al., 2015; reviewed in Feldman, 2015; Hasson & Frith, 2016; Hasson et al., 2012; Schoot, Hagoort, & Segaert, 2016; Stolk, Verhagen, & Toni, 2016). Indeed, the quality and amount of action-perception couplings over the course of early development better predicts later language ability (Hirsh-Pasek et al., 2015) and language-related brain function (Romeo et al., 2018) than more traditional measures, such as the number of words heard (Lindsay, Gambi, & Rabagliati, 2019). Similarly, synchronous interbrain (limbic) dynamics in early infancy (i.e., prior to the onset of cooperative pointing) appears to be concomitant with several kinds of positive social experience, such as closeness and social bonding[12] (Atzil, Hendler, & Feldman, 2014; Atzil et al., 2017; reviewed in Feldman, 2015; 2017).

**Dynamics at the timescale of ontogeny**

The dynamics sketched in Section 4.1.2 suggests a kind of Vygotskian scaffolding (Vygotsky, 1978) or 'co-construction' (Tomasello, 2019) of the dynamics of internal states; whereby – via recurrent engagement in loops of coupled action-perception with relatively

---

[12] To be clear, our claim is *not* that *only* the usage of communicative constructions can give rise to interbrain synchrony (see, e.g., evidence of nonverbal interbrain synchrony and associations with feelings of interpersonal closeness, Kinreich, Djalovski, Kraus, Louzoun, & Feldman, 2017). Communicative constructions are merely a (highly useful) means to gather reliable evidence for the adaptive prior specified in the main text.

'entrenched' aspects of the niche – individuals learn (internalize) the salience of culturally anticipated policies used to infer hidden states. That is, by acting in a shared environment that contains older, relatively inflexible individuals that perform stereotyped behavior (characteristic of 'how we do things here'), younger individuals are able to learn the deontic value of policies (Ramstead et al., 2016; Veissière et al., 2019). For our purposes, this means that individuals' prior beliefs become more similar across couplings through (bidirectional) processes of (asymmetric) enculturation[13] (Renzi, Romberg, Bolger, & Newman, 2017). That is, recurrent episodes of acutely increased alignment – of the kind typical of coupled action-perception cycles – are necessary for the creation and maintenance of species-typical states. In short, to gather evidence for an adaptive prior that mental states are aligned, one must act to bring about sensory states that are indicative of this belief (Byrge, Sporns, & Smith, 2014; Chiel & Beer, 1997).

Within and across interactions, such a dynamics increases the adaptive value of, e.g., collaborative foraging strategies by increasing inferred reliability in the hidden states generating observations (others' intentions; Han, Santos, Lenaerts, & Pereira, 2015; Nakamura & Ohtsuki, 2016). This is because gathering evidence for the prior beliefs of other agents entails predicting how their beliefs relate to the niche; i.e., how others' beliefs relate to one's own mental states as well as non-social affordances. Consequently, gathering reliable evidence for others' mental states entails redirecting attention triadically (jointly). In this way, individuals become more reliable models of their interlocutor(s), and hence may leverage their own expectations about others' actions to guide expectations over sensory outcomes, like couplings with environmental affordances[14] (e.g., Bach, Nicholson, & Hudson, 2014; Gallotti & Frith, 2013; Pezzulo, 2011).

---

[13] For expository purposes we leave undiscussed the ontogenesis of critically important and later-appearing interactions with peers (e.g., Ashley & Tomasello, 1998; Brownell & Carriger, 1990; Brownell, Ramani, & Zerwas, 2006; see Brownell & The Early Social Development Research Lab, 2016). Future explorations leveraging this approach should look to integrate data relating to peer-peer interactions in ontogenesis (reviewed in Brownell, 2011). Indeed, such phenomena are of great interest to the present account given the (possibly) more complex dynamics exhibited by the attunement of two systems embodying relatively imprecise expectations about how best to minimise uncertainty (e.g., Eckerman, Davis, & Didow, 1989).

A useful way to increase the degree of alignment of prior beliefs among individuals is to send more information to one's communicative partner. Holding the inferred common ground constant, one of the main ways to convey more information is to allow for hierarchically deeper policies (e.g., sequences of sequences) to generate action; that is, roughly, to provide more form (i.e., use longer communicative constructions). In effect, more information about mental states is thereby made observable. This perspective sheds interesting light on the species-typical trajectory from triadic attention (Striano & Stahl, 2005) to more reliably enacted forms of joint attention underwritten by reciprocal information flow – and the usage of pointing and gesture (Carpenter & Liebal, 2011; Tomasello et al., 2007) – to more complex constructions leveraged to transact with the hidden mental states of others (Aureli & Presaghi, 2010; see Colonnesi, Stams, Koster, & Noom, 2010). The human agent appears to build up, nuance, and consolidate its (mutually expected) repertoire of action policies that, based on experience, have proven useful for adequately attuning with the mental states of conspecifics. That is, through this kind of continuous growth and hierarchical differentiation in communicative action policies (Goldin-Meadow, 2007; Tomasello, 2008), human individuals appear as though they were learning to tune themselves to the niche, and the niche to themselves.

Speaking generally, by repeatedly engaging in coupled action-perception cycles, individuals distil and abstract deeper observation-policy mappings (i.e., constructions) from the bottom up; that is, on an item-by-item basis (reviewed in Tomasello, 2000). In certain cases, individuals may then leverage learned hypotheses (about how best to disambiguate mental states) to reliably constrain the hypothesis space for learning and inference about constructions[15] (McClelland et al., 2010; see also Tenenbaum, Kemp, Griffiths, & Goodman, 2011). That is, induction at higher layers of the model can serve to bootstrap learning at lower layers. Such 'domain-general' learning processes are illustrated by the model of Perfors, Tenenbaum, & Regier (2011). These authors provide a proof of principle account showing that several hours of child-directed input is sufficient for the posterior expectations of a hierarchical approximate Bayesian (i.e., active inference) learner – leveraging domain-general learning mechanisms – to converge towards a single, high level hypothesis about the causes of sensory input (here, a set of context-free grammars). That is, this set of context-free grammars had the greatest probability at the end of training. Consequently, this empirical prior functioned as abstract knowledge – it constrained expectations about likely hypotheses (in particular, auxiliary fronting) at lower layers[16] (see also Kemp, Perfors, & Tenenbaum, 2007).

---

[14] From this perspective, it may be interesting for modelling work to investigate the notion of joint attention as an emergent property of coupling two hierarchical generative models attempting to infer the hidden states of each other (e.g., Friston & Frith, 2015a) while embedded in a broader ecological niche (e.g., Williams & Yaeger, 2017). In such a context, does joint attention effectively function to minimise a sensory Lagrangian over (jointly anticipated) sensory states (Sengupta et al., 2016)? What role does cooperative communication play in maintaining such a gauge invariance over the action of shared sensory states?

Indeed, modelling schemes employing active inference provide evidence of their utility for modelling attunement to a communicative system (e.g., Friston, Rosch, Parr, Price, & Bowman, 2017; Kiebel, Daunizeau, & Friston, 2008; Kiebel, von Kriegstein, Daunizeau, & Friston, 2009).

---

[15] Relatedly, because higher, contextualising layers of a hierarchical model sample a larger space of inputs in estimating a smaller number of (more abstract) hypotheses (Tenenbaum et al., 2011), agents may, in certain instances, learn contextualising 'overhypotheses' faster than learning at lower layers of the model (Gershman, 2017).

[16] We urge the reader to take the model of Perfors, Tenenbaum, & Regier (2011) with some caution. This is because part of the specification of their model was a set of (sets of) grammars; that is, their model came 'pre-equipped' with knowledge of various (formal, arbitrary) grammatical 'principles'. Thus, their model had simply to converge on the most probable set of grammars (hypothesis) given its 'innate' repertoire. However, there is i) no clear evidence for such innately specified (i.e., formal and arbitrary) linguistic principles in humans (Dąbrowska, 2015); ii) no clear formulation of what may be included in such an innate repertoire (Tomasello, 2004); and iii) numerous logical problems with the evolution of such innate structure (Christiansen & Chater, 2008; cf. e.g., Berwick, Friederici, Chomsky, & Bolhuis, 2013).

For instance, Yildiz, von Kriegstein, & Kiebel (2013) used the active inference formalism to model word learning under optimal and noisy conditions and under variations in speaker accent. By attending to incoming input (i.e., increasing the precision of sensory signals), their model tuned its top-down beliefs to the structure of training data, which comprised sequences (of sequences) of spoken phonemes. The authors report that this model outperformed other computational learning schemes across a range of conditions and could be used to explain the judgements of adult second language learners. Future modelling work should investigate how an adaptive prior for alignment covers more ecologically valid instances of attunement to communicative constructions, such as the effects that 'starting small' and a prolonged period of developmental immaturity have on attuning to a communicative system (Bjorklund, 1997; Elman, 1993).

As noted, alignment with communicative partners means learning a set of 'automatic,' experientially robust (deontic) observation-policy mappings; e.g., the expectation (for English speakers) that a determiner typically precedes a noun (Meylan, Frank, Roy, & Levy, 2017). Indeed, this view fits nicely with usage-based approaches to language acquisition (Lieven, 2016; Tomasello, 2003). Proponents of this view suggest that "constructions of all types are automatized motor routines and subroutines" that "come out of language use in context and… cognitive skills and strategies used in non-linguistic tasks" (Bybee, 2003, both p. 158).

Indeed, much of the structure and dynamics of the neural regions that underwrite the learning and usage of cooperative communication have been exapted (in particular, 'cooperativized') from their earlier evolutionary functions (Anderson, 2010; Bjorklund & Ellis, 2014). This has been emphasized, for instance, by embodied neurosemantics models of the neural underpinnings of the acquisition and comprehension of meaning in form-meaning pairings (reviewed in Pulvermüller, 2013; 2015). In such approaches, the meanings of both concrete and abstract constructions (e.g., 'kick' and 'love,' respectively) are grounded in low level sensorimotor dynamics and action-perception circuits contextualized by top-down input (Moseley et al., 2012; also, Harnad, 1990).

To elaborate, human brains effectively combine two kinds – or two hierarchical levels – of general-purpose learning architecture to capitalize on the epistemic opportunities afforded by the action-perception cycle: (i) self-supervised (approximate Bayesian) learning, via the dynamic, hierarchical interplay between descending, neuronally encoded predictions and ascending prediction errors over time (Badcock et al., 2019b); and (ii) supervised (social) learning in a cultural niche via repeated, immersive practice in a set of culturally patterned routines (Ramstead et al., 2016; Roepstorff, Niewöhner, & Beck, 2010). In effect, attuning to a system of communication constructions requires learning how to process and use form-meaning pairings in real-time communication. Thus, we are in agreement with Christiansen & Chater (2016a), who note that "learning [a cooperative communication system] involves creating a predictive model of the language, using online error-driven learning" (p. 121).

Deploying these learning processes in species typical communicative couplings means that, on average over time, individuals' communicative action policies become sufficiently similar; that is, not identical, but usable (Kidd, Donnelly, & Christiansen, 2018; Tomasello, 2003). This is depicted in Figure 6. For instance, Bannard, Lieven, and Tomasello (2009) found that the perplexity (an information theoretic measure that quantifies the fit of a distribution to a set of observations) of the (probabilistic) context-free grammar used to capture one child's (Brian's) utterances at age 2;0 was able to account for approximately 15% of the utterances of another child (Annie, also 2;0). Similarly, the grammar imputed to Annie at 2;0 was able to explain approximately 36% of the utterances for Brian (2;0). Interestingly, at 3;0 model fit in either direction was increased. Thus, the grammar imputed to Brian at 3;0 accounted for roughly 59% of Annie's utterances (3;0), while the grammar imputed to Annie at 3;0 accounted for about 63% of Brian's utterances (3;0). Though the authors did not compute the significance of this change, this trend is precisely what one would expect under our model; namely, a trend towards statistically similar prior beliefs over hidden causes as individuals converge towards their cultural attractor (Figure 6).

In sum, by repeatedly 'filtering' one's action through others' mental states, one obtains a useful set of policies for flexibly and economically disambiguating prior beliefs. These correspond to policies with a high deontic value (Constant et al., 2019). In this way, one's set of constructions appears to converge on the set of constructions that constitute the communicative system(s) that predominantly generate one's sensory samples (i.e., those used by one's speaker community). This is to say that the prior beliefs of individuals converge towards an exploitable degree of similarity. One thus instantiates a sufficiently reliable model of the processes that generate sensory observations. This ontogenetic tendency repeats itself, cyclically, across generations. This has critical implications for the historical development of least effort communicative systems, to which we now turn.

**Dynamics at the timescale of cultural evolution**

According to the model of cooperative communication proposed here, communicative systems (i.e., the dynamics of sets of form-meaning pairings) should appear, on average across time, to minimize their variational free energy (Ramstead et al., 2018). But what, exactly, does this mean; and how might this claim be investigated empirically? As noted above, a pointing gesture does not, in general, allow an agent to infer hidden causes as efficiently or reliably as the usage of a more complex construction. Therefore, in attempting to minimize their free energy, communicative systems should evolve towards a balance between usability and learnability (simplicity) on the one hand, and on the other, increasingly arbitrary, hierarchically deeper (complex) action sequences. This means that communicative systems should appear to optimize an accuracy-complexity, or expressivity-compressibility, trade-off (Tamariz & Kirby, 2016). Consider, for instance, the "drift to the arbitrary" proposed by Tomasello (2008, p. 219). Here, the suggestion is that 'grainy' bodily gestures like pointing give way to increasingly expressive, 'finer grained' gestures like pantomime. In turn, relatively expressive gestures give way to even more expressive, 'finely grained' vocal gestures like abstract and arbitrary communicative constructions (for similar views, see Fay, Arbib, & Garrod, 2013; Perniss & Vigliocco, 2014; Wilcox, 2004). Interestingly, this roughly recapitulates the general ontogenetic trajectory of cooperative communication described above. It is as though – at multiple, nested scales of analysis – the human agent were becoming increasingly adept at flexibly deploying an increasingly sophisticated set of actions to resolve its sensory ambiguity.

These ideas are supported by the finding that the dynamics of relevance optimisation across recurrent interactions, and generations of speakers, manifests in constructions that increase in expressivity with respect to production and processing costs (Tamariz & Kirby, 2016; e.g., Kirby, Tamariz, Cornish, & Smith, 2015; Fay et al., 2010). This means that human communicative systems, after a sufficiently long period of evolution, tend to cluster in a kind of 'least effort' subregion of a design space (i.e., parameter space) of communicative systems[17] (i Cancho & Solé, 2003; Seoane & Solé, 2018; see Dediu et al., 2013; Evans & Levinson, 2009). Expressed otherwise, relative to earlier generations of users of a particular communicative system, individuals in subsequent generations may be advantaged with respect to the range of communicative constructions that can be used to disambiguate mental states (Angus & Newton, 2015; perhaps, e.g., by coming to distinguish among previously undistinguished actions; Senghas, 2003). That is, communicative constructions themselves evolve to 'fit,' or gather evidence for, the adaptive priors favored by evolution and the specific demands of the local ecological niche (Christiansen & Chater, 2008; Kirby, Dowman, & Griffiths, 2007; Perfors & Navarro, 2014). This means that, over historical time, processes of cumulative cultural evolution (e.g., Henrich, 2015; Richerson & Boyd, 2005) tend to increase the deontic value of constructions by increasing the expressivity and complexity of using and learning such constructions (for similar viewpoints, see Christiansen & Chater, 2016a; Cornish, Tamariz, & Kirby, 2009; Dingemanse, Blasi, Lupyan, Christiansen, & Monaghan, 2015; Fay et al., 2018; Kirby et al., 2015; Tamariz & Kirby, 2016).

---

[17] To be clear, we are by no means claiming that this set of parameters exists in the brains of speakers. We are talking here about a scheme for modeling the dynamics of the cultural evolution of human communicative systems.

Cooperative communication emerges as a multiscale, self-organizing process that unfolds simultaneously across interaction, ontogeny, and cultural evolution (also, de Boer, 2011). Consequently, the adaptive prior under consideration enables, drives, and sustains each scale of dynamics. Circularly, each scale of dynamics generates actions that appear to gather evidence for the adaptive prior. Across developmental time, the contextualizing dynamics of cultural evolution appear as a higher-order attractor – itself evolving in time, but sufficiently stable from the perspective of the developing individual – towards which individuals converge via recurrent engagement in coupled action-perception cycles that unfold in real-time. Taken together, interlocked dynamics at these three scales entrench the existence (i.e., probability) of the adaptive prior. In this way, cooperative communication becomes a self-fulfilling prophecy. That is, by gathering evidence for their adaptive priors, the low-level dynamics of interactants appear to create and maintain, at least for some period, the observable coherence of a contextualizing scale of (cultural) organization; namely, a communicative system (also see Szathmáry, 2015).

In active inference, the partitioning in the timescales that characterize a communicative system is formalized as between-scale differences in the precision of prior beliefs as one ascends scales (Constant et al., 2018; Ramstead et al., 2018). This is the result of, e.g., increasing the number of components (Smith et al., 2017) and the connectivity between components constituting a communicative system (Reali, Chater, & Christiansen, 2018). This means that linear modifications to inputs to the system are associated with nonlinear changes in its dynamics (Beckner et al., 2009; Shuai & Gong, 2014). Nonlinearity is an inherent property of self-organizing systems (Prigogine & Stengers, 1984) and manifests in phenomena like critical slowing (i.e., phase transition; Gandhi, Levin, & Orszag, 1998; i Cancho & Solé, 2003), parameter reduction (Riley, Richardson, Shockley, & Ramenzoni, 2011), and chaotic dynamics (Sanders, Farmer, & Galla, 2018). A change in the characteristic timescale of the dynamics of a cooperative communicative system is exemplified by Smith et al. (2017). These authors report experimental and simulation results suggesting that multi-person communicative systems exhibit slower regularization (decrease in conditional entropy) of a plurality marker across generations relative to communicative systems constituted by a single individual (for discussion of disparities of the pace of change across communicative systems, see Gray, Greenhill, & Atkinson, 2013).

As noted above, the evolution of a communicative system may be cast as motion through a design space of communicative systems. Such spaces are effectively equivalent to the linguistic morphospace (Gray et al., 2013), or the space of states taken on by human communicative systems (e.g., linguistic networks; Seoane and Solé, 2018). Motion in design space may be relatively simple. For instance, Bybee (2010) has suggested that processes of grammaticalization – where flexible lexical forms gradually transition to fixed grammatical forms – may be modelled in terms of unidirectional (i.e., irreversible) motion through a continuous parameter space (also see Haspelmath, 1999). This might be modelled as a strange (Lorentz) attractor (Bybee, 2010), similar to that observed in models of communicative alignment (Friston & Frith, 2015a). In some cases, this motion may be more complex. For instance, the selection pressures acting on a system's constructions and, hence, the evolutionary trajectory of that set of constructions, varies as a function of the size of the population of speakers (Fay & Ellison, 2013; Lupyan & Dale, 2010; Reali et al., 2018; see Dingemanse et al., 2015).

In sum, the cultural niche construction implicit in free energy minimization in an ensemble of communicating conspecifics can be seen as a form of active inference on a (cultural) evolutionary level. In other words, selection pressures are just free energy gradients that allow us to cast *selection* (for useful communicative constructions) as a process of Bayesian *model selection* to maximize fitness; i.e., model evidence or the probability of communicative exchange, under a shared generative (phenotypic) model. This perspective nicely combines structure learning, evolution, and niche construction within the same formalism. For further discussion, please see Campbell (2016), Constant et al. (2018), Frank (2012), and Sella and Hirsh (2005).

## Future Directions and Conclusion

In this article, we have outlined an extension to existing theories of cooperative communication. Our extension is based on active inference and provides a novel, integrative take on the biobehavioral underpinnings of cooperative communication that complements existing psychological accounts (Tomasello, 2003; 2008; 2014; 2019). A more complete account of the dynamics entailed by the adaptive prior for alignment requires an integrative approach to research. To be sufficient, such research must aim to encapsulate the various timescales from which this prior emerges, particularly in a way that renders each scale of analysis complementary and mutually constraining with respect to the others (Badcock et al., 2019b; Ramstead et al., 2018; Tinbergen, 1963). The initial, though surely not exclusive, timescales of interest for cooperative communication were outlined in this paper. These range from the evolutionary history of early humans, to the intergenerational transmission of cultural patterns, down to individual development, and to two people conversing in real-time. cultural patterns, down to individual development, and to two people conversing in real-time. This multiscale framework, arising from and underwriting the dynamics of the adaptive prior for alignment, should help to facilitate an understanding of inter- and intracultural similarities and differences in the structure and function of culture, mind, and brain. We conclude with a few comments about the limitations of the current proposal for the adaptive prior for alignment.

One limitation is the relative dearth of 'direct' evidence generated by empirical and computational studies of cooperative communication – guided by the notion of an adaptive prior for alignment. We admit this is an important weakness, although one which can only be remedied through future research. Nevertheless, we have reviewed a substantial amount of indirect evidence generated by a range of empirical and simulation studies that speak to the integrative potential of the adaptive prior for alignment in making sense of cooperative communication.

For instance, our approach can be used to model the neuronal message-passing underwriting cooperative communication, as implied by active inference (e.g., Bastos et al., 2012; Parr & Friston, 2018). To illustrate this, regions in higher layers of cortex, such as anterior cingulate cortex (ACC; van den Heuvel & Sporns, 2013), integrate limbic afferents encoding salience with control policies issued by motor cortex (Friston et al., 2014b; Pezzulo et al., 2018). In turn, descending connections from paralimbic cortex convey signals that are unpacked as hierarchically nested sequences of cooperatively motivated action and inference, such as declarative pointing (Brunetti et al., 2014; see Apps, Rushworth, & Chang, 2016; Chambon et al., 2017; Haroush & Williams, 2015; Holroyd & Yeung, 2012; Lavin et al., 2013). Interestingly, these neural considerations align with the psychological suggestions of Hare and Tomasello (2005), who suggest that early human selection pressures favored novel limbic dynamics that encode an increased tolerance and trust for conspecifics in the context of food (see the self-domestication hypothesis; Hare, 2017).

One specific, promising modelling approach pertains the usage of hierarchies of stable heteroclinic channels (SHCs; e.g., Rabinovich, Varona, Tristan, & Afraimovich, 2014). SHCs are neuronally plausible models of hierarchically deep sequences (i.e., state trajectories of state trajectories) that may be scaled up to account for the acquisition and processing of more realistic cooperative communication data than have thus far been examined (Kiebel et al., 2009; see Rabinovich, Simmons, & Varona, 2015). In particular, it may be possible to use such a scheme to model the processing and use of communicative constructions, as these are hierarchically deep sequences of sequences (i.e., constructions are a statistically reliable ordering or 'chunk' of, e.g., word classes that entail chunks of morphemes that entail chunks of phonemes; relatedly, see 'chunk and pass' processing; Christiansen & Chater, 2016b). Indeed, the (re)use of hierarchical processing for language use may represent one instance of cooperativized, domain-general cognition exapted for usage in a cooperative social milieu. This is evidenced, for instance, by the presence of hierarchical processing of an artificial communication system in infants before nine months of age (Kovács & Endress, 2014). Such a (developing) processing capacity may then be biased by the adaptive prior for alignment, after nine to twelve months of age, towards disambiguating hierarchically organized communicative constructions (see Elman, 1993).

Another limitation of our proposal is that our consideration of the ontogenetic trajectory of cooperative communication focused exclusively on its typical trajectory. This was due to concerns about space. We readily acknowledge that there are all kinds of species atypical (i.e., unexpected) trajectories for the phenotypic expression of the adaptive prior for alignment (indeed, this is one of the reasons it is so interesting to study!). Arguably, studying how the dynamics of the adaptive prior for alignment may be perturbed in ontogenesis is crucial (e.g., discerning neurocomputational atypicalities or atypicalities in local niche dynamics; Thomas et al., 2019). Gaining a fuller grasp on the adaptive prior for alignment requires the integration of data and theory not just '*vertically*' (i.e., across scales), but also '*horizontally*' between the niche and its denizens. That is, the adaptive prior for alignment manifests distinctively not only across an array of timescales, but also at any given time across an array of cultural settings and, within cultures, neurotypical and neurodiverse populations.

For instance, in section 4, we discussed how, in neurotypical individuals, adequately explaining away sensory causes depends on a delicate, finely tuned balance of the top-down precision of hypotheses and the bottom-up precision of sensory fluctuations. But consider the case of autism, where neurocomputational atypicalities are thought to render the individual oversensitive to incoming error signals (Lawson, Rees, & Friston, 2014; Mirza, Adams, Friston, & Parr, 2019; see also Thomas et al., 2016). Such individuals would still expect to align mental states (Jaswal & Akhtar, 2019), and so would attend to others' communicative behaviors, but would be unable to attenuate the precision of sensory signals (Hadjikhani et al., 2017; Mirza et al., 2019).

Consequently, during initial interactive couplings, such individuals might initially look like they are typically developing (i.e., attending to others' eye gaze; Young, Marin, Rogers, & Ozonoff, 2009). However, repeated attention to sources of sensory uncertainty (e.g., others' saccades), combined with an inability to adequately leverage predictions to explain away this uncertainty (owing to too much sensory precision), means that such individuals may develop idiosyncratic or atypical phenotypic expressions of the adaptive prior for alignment (e.g., avoiding eye gaze; Tanaka & Sung, 2015). In other words, early atypicalities in the internal dynamics generating evidence gathering cycles of action-perception may have downstream effects on joint attentional skills (Charman et al., 1997; Nyström et al., 2019), attunement to and use of communicative action policies (Loveland & Landry, 1986; Warlaumont, Richards, Gilkerson, & Oller, 2014), mental state inference (Tager-Flusberg, 2007), and other means for alignment (Heasman & Gillespie, 2019). In short, aberrant inference in a prosocial, developmental setting may easily lead to a pernicious kind of dyslexia – not for the written word – but for any communicative exchange (i.e., joint inference).

In summary, the adaptive prior for alignment 'sets the tone,' as it were, for species-typical patterns of evidence gathering, about oneself and the (social) world, that unfold over different timescales. The adaptive prior for alignment is, in effect, a kind of 'best guess' about the state occupied by the system at any point in time. Thus, for a human, processes of action, inference, learning, and (cultural) niche construction appear as if they were, on average across time, in the service of gathering evidence for the hypothesis that 'I' am like 'you', that 'you' are like 'me', and that 'we' exist.

## Acknowledgements

We thank Michael J. Farrar, Robert D'Amico, Andreas Keil, Michael Tomasello, Leon Li, and Josh Perlin for helpful commentary, discussion, and feedback on early drafts of the manuscript. This research was produced thanks in part to funding from the Canada First Research Excellence Fund, awarded to McGill University for the Healthy Brains for Healthy Lives initiative (S. P. L. Veissière and M. J. D. Ramstead), as well as a Postgraduate Research Scholarship in the Philosophy of Biomedicine as part of the ARC Australian Laureate Fellowship project A Philosophy of Medicine for the 21st Century (Ref: FL170100160) (A. Constant), a Social Sciences and Humanities Research Council doctoral fellowship (Ref: 752-2019-0065) (A. Constant), a Joseph-Armand Bombardier Canada Doctoral Scholarship and a Michael Smith Foreign Study Supplements award from the Social Sciences and Humanities Research Council of Canada (M. J. D. Ramstead), and by a Wellcome Principal Research Fellowship (K. J. Friston – Ref: 088130/Z/09/Z).

## Author Contributions Statement

JV and MJDR conceptualized the work for the paper and JV identified the paper's main target, i.e., cooperative communication. JV, PB, and MJDR worked out the argument and planned the paper. JV wrote the first draft of the manuscript. PB and AC helped with conceptualization work. AC drafted parts of section 3. PB helped edit the paper. KF and MJDR ensured that the technical aspects of the paper were presented accurately.

## Conflict of Interest Statement

The authors declare no conflicts of interest.

Submitted manuscript (post-peer-review, pre-copyedit). Frontiers in Psychology – Cultural Psychology. Special Issue: Imagining Culture Science: New Directions and Provocations. Please do not cite this version.

*Research in Child Development*, *63*(4), i – vi, 1–143.

Chambon, V., Domenech, P., Jacquet, P. O., Barbalat, G., Bouton, S., Pacherie, E., … Farrer, C. (2017). Neural coding of prior expectations in hierarchical intention inference. *Scientific Reports*, *7*(1), 1278. https://doi.org/10.1038/s41598-017-01414-y

Charman, T., Swettenham, J., Baron-Cohen, S., Cox, A., Baird, G., & Drew, A. (1997). Infants with autism: An investigation of empathy, pretend play, joint attention, and imitation. *Developmental Psychology*, *33*(5), 781–789. https://doi.org/10.1037//0012-1649.33.5.781

Chiel, H. J., & Beer, R. D. (1997). The brain has a body: adaptive behavior emerges from interactions of nervous system, body and environment. *Trends in Neurosciences*, *20*(12), 553–557. Retrieved from https://www.ncbi.nlm.nih.gov/pubmed/9416664

Christiansen, M. H., & Chater, N. (2008). Language as shaped by the brain. *The Behavioral and Brain Sciences*, *31*(5), 489–508; discussion 509–558. https://doi.org/10.1017/S0140525X08004998

Christiansen, M. H., & Chater, N. (2016a). *Creating Language: Integrating Evolution, Acquisition, and Processing*. Cambridge, MA: The MIT Press.

Christiansen, M. H., & Chater, N. (2016b). The Now-or-Never bottleneck: A fundamental constraint on language. *Behavioral and Brain Sciences*, *39*, e62. https://doi.org/10.1017/S0140525X1500031X

Christiansen, M. H., & Kirby, S. (2003). Language evolution: consensus and controversies. *Trends in Cognitive Sciences*, *7*(7), 300–307. Retrieved from https://www.ncbi.nlm.nih.gov/pubmed/12860188

Chukoskie, L., Snider, J., Mozer, M. C., Krauzlis, R. J., & Sejnowski, T. J. (2013). Learning where to look for a hidden target. *Proceedings of the National Academy of Sciences of the*

**Figure 1.**

| Scale of Analysis | Characteristic Dynamics and Processes of Cooperative Communication |
|---|---|
| Real-Time (Interaction) | • Ostension<br>• Joint attention<br>• Relevance optimization<br>• Coupled, bidirectional flow of information<br>• Proximate motivation to align and coordinate mental states (e.g., declarative, interrogative, and informative motives) |
| Development (Ontogeny) | • Clearest behavioral onset at ~9-12 months of age (i.e., cooperative pointing)<br>• Gradual alignment with a conventionalized communicative system |
| Cultural Evolution (Phylogeny) | • Historical development of a communicative system (e.g., grammaticalization, syntactic reanalysis, semantic bleaching)<br>• Diversification of communicative systems across time, space, and speaker communities |
| Biological Evolution (Evolution) | • 'Cooperativization' of non-human great ape communicative motives and skills (e.g., non-human ape attention getters as precursor to human declarative pointing)<br>• Ultimate motivation to align and coordinate mental states (i.e., mutual expectations of cooperativeness) |

**Fig. 1**. *Summary of key features circumscribing cooperative communication.* Certain features, e.g., alignment with a communicative system, are discussed in Section 4. This paper associates these phenomena with the corresponding scale of dynamics underwritten by the free-energy formulation and, more substantively, the hierarchically mechanistic mind (see Fig. 4 in Badcock, Friston, & Ramstead, 2019).

**Figure 2.**

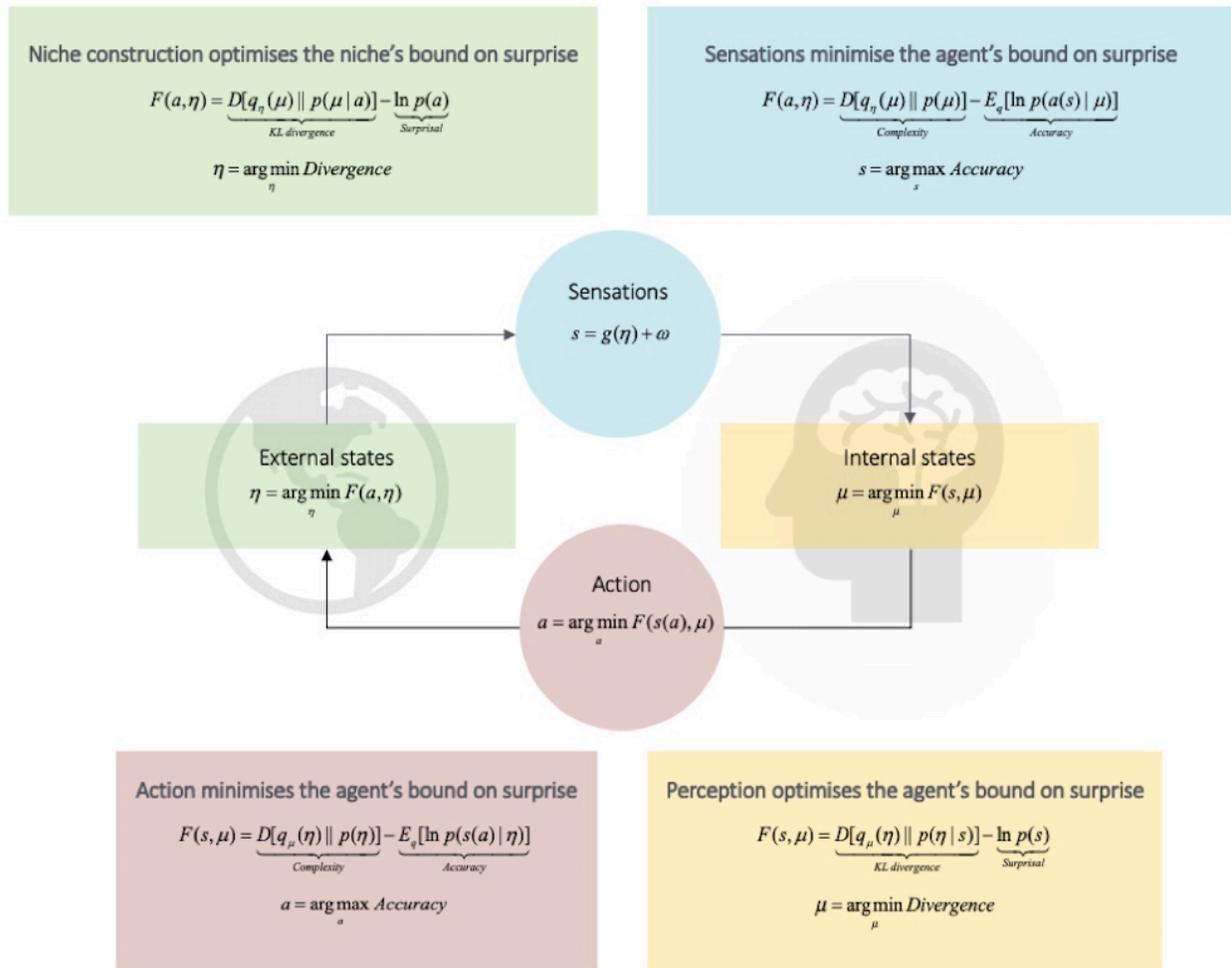

Fig. 2. *Active inference.* This Figure schematizes active inference. It depicts the coupling of an agent's internal states (the dynamics of which entail predictions or beliefs about the niche, $\mu$) to its external states (the dynamics of the agent's niche, $\eta$). Middle Panel: The influence of the niche on the agent is given by the dynamics of the agent's sensations, $s$. Reciprocally, the influence of the agent upon its niche is given by the agent's action, $a$, upon the niche. This means that the niche is not directly observable from the perspective of an agent's internal states; and the agent's internal states are not directly observable from the perspective of the niche. From an agent's

perspective, the niche is thus described as a set of *hidden variables*. Hidden variables must be inferred (i.e., predicted) from sensory observations. Thus, to minimize the probability of sampling surprising sensory states, the task for the agent is to attune the dynamics of internal states to those of the niche; or attune the dynamics of the niche to those of internal states. Attunement renders the agent an approximate (predictive) model of the hidden causes of its sensations. We can quantify the degree of attunement between organism and niche with a quantity called variational free energy (Bruineberg et al., 2018; Constant et al., 2018). Free energy $F$ bounds (i.e., is greater or equal to) the *surprisal $-ln\ p(s)$* associated with a sensation (Friston, 2010). Importantly, free energy is a function of two quantities to which the organism has access, namely, its sensations and predictions (for discussion, see Bruineberg et al., 2018). Lower Panel: The bottom right details how perception optimizes free energy by implicitly minimizing a Kullback-Leibler (KL) divergence term $D$. The KL divergence tracks the statistical similarity of two distributions (Cover and Thomas, 1991); e.g., the similarity of prior beliefs about the state of the niche with posterior beliefs (Friston, 2012). Because the KL divergence provides an upper bound on surprisal, minimizing it renders the agent a model of the niche and thus implicitly bounds the surprise of sensory states. Upper panel: These expressions define the relationship of the niche to the agent. Note the kind of 'mirror image' relationship between the equations in the upper panel with the equations in the lower. This relationship is a consequence of the mathematics of free energy minimization (see Bruineberg, Rietveld, et al., 2018; Constant et al., 2018). It means that the niche 'sees' and 'learns' about the agent (i.e., via the agent's action) in the same way an agent sees and learns about their niche (i.e., via the niche's 'action'). This insight is extended in Fig 3. Adapted with permission from Veissière et al. (2019).

**Figure 3.**

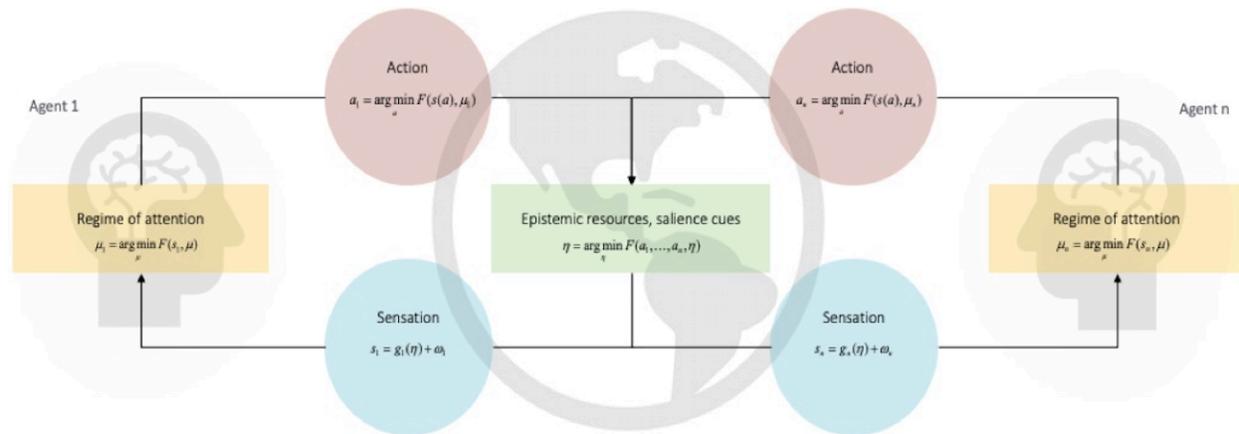

**Fig. 3**. *Thinking through other minds.* This Figure depicts a set of heuristic equations that describe the kind of free energy minimization hypothesized to underwrite the acquisition and production of learned cultural behaviors via the coupled dynamics sketched in the main text (full equations in Figure 2). In the context of human communication, coupled dynamics are energized by an adaptive prior for alignment. The adaptive prior for alignment specifies the characteristically enhanced precision of the hypothesis that 'we' exist. This prior motivates similar agents to actively couple their respective actions $a_n$ and sensations $s_n$. Via the processes discussed in the main text, this statistical coupling of sensation and action enables each individual to reliably to align with (i.e., to infer) the hidden states $\mu_n$ of conspecific $n$. This circular process brings about a process of cultural niche construction that creates, maintains, and modifies a set of predictable epistemic (i.e., deontic) resources, $\eta$. These specify a set of high value (i.e., predictable) observation-policy mappings, which are used to disambiguate the mental states of conspecifics (Veissière et al., 2019). One important class of deontic resource is the set of observation-policy mappings that underwrite a system of communicative constructions (i.e., form-meaning pairings). This means that the use of communicative constructions plays a critical role in enabling agents with an adaptive prior for alignment to effectively disambiguate external states. This is because an agent's external states are constituted, in part, by the internal, mental states of another agent (and vice versa). This follows form the fact that external states cause sensation; for an agent equipped with an adaptive prior for alignment, inferring the motion of external states entails inferring other agents' hidden states. The production and observation of communicative constructions is useful because it effectively and flexibly guides 'regimes' of attention that enable species' unique forms of cultural learning (see Ramstead, Veissière, & Kirmayer, 2016; Veissière et al., 2019). Diachronically, communicative constructions are finessed by a community of agents via the inheritance and (intended or unintended) modification of constructions during either learning or usage. Adapted with permission from Veissière et al. (2019).

**Figure 4.**

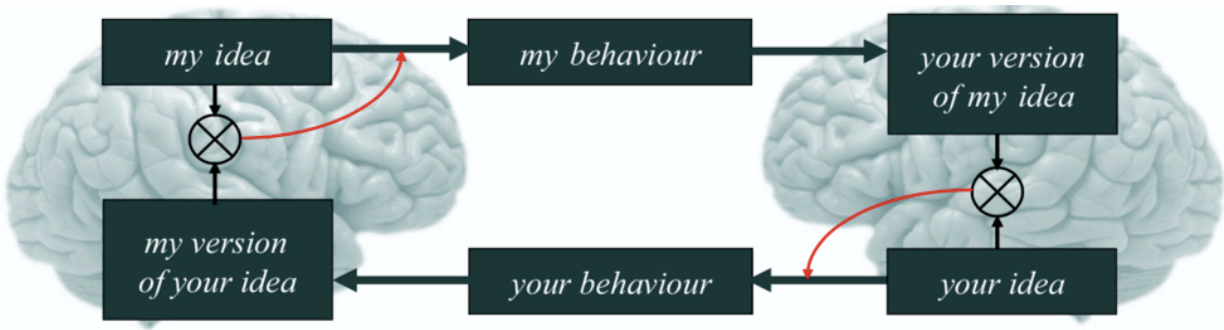

**Fig. 4.** *One canonical 'loop' of the coupled action-perception cycle.* This example is 'canonical' in the sense that the manifestation of the coupled action-perception cycle in a given instance may vary as a function of context and the experience of its constituent members (e.g., an infant's communicative needs with an adult are different than a pair of adults'). With this in mind, for two agents A and B expecting to reliably infer each other's mental states, the beliefs of A ('my idea') generate A's observable actions ('my behavior'). The actions of A, in turn, cause the (attended) sensory states of B. Attention directed towards agent A by agent B in turn enables the observations generated by A to entrain the hidden states of B ('your version of my idea'). This is just some hypothesis entertained by B about the causes of B's observations (i.e., about the mental states generating A's actions). To increase or maintain the reliability of B's hypothesis, B must then act on the niche ('your behavior') to test B's hypothesis about hidden causes, as it were (that is, to check for mutual understanding, for instance; e.g., Clark & Wilkes-Gibbs, 1986). B thereby causes A's attended observations and, hence, A's mental states ('my version of your idea'). This looping dynamics continues until both agents infer alignment (Friston & Frith, 2015a). Central here is that A is attending to the sensory states generated by B (and *vice versa*) because the only way to gather evidence for the adaptive prior that mental states are aligned is to attend to the sensory effects of one's actions; and evidence for hypotheses about the sensory effects of one's actions can *only* be given (in the present context) by the actions of the other agent. Working backwards, because the actions of another agent are generated by their mental states; and their mental states are entrained by (attended) sensory observations; and their sensory observations are generated by one's own actions; we thus arrive at the claim, given at the start of this section, that "A central part of the content of the prior belief prescribing the alignment of mental states among conspecifics is that the actions of agents (e.g., oneself) modulate the mental states (prior beliefs) of other agents."

Adapted with permission from Friston & Frith (2015a).

**Figure 5.**

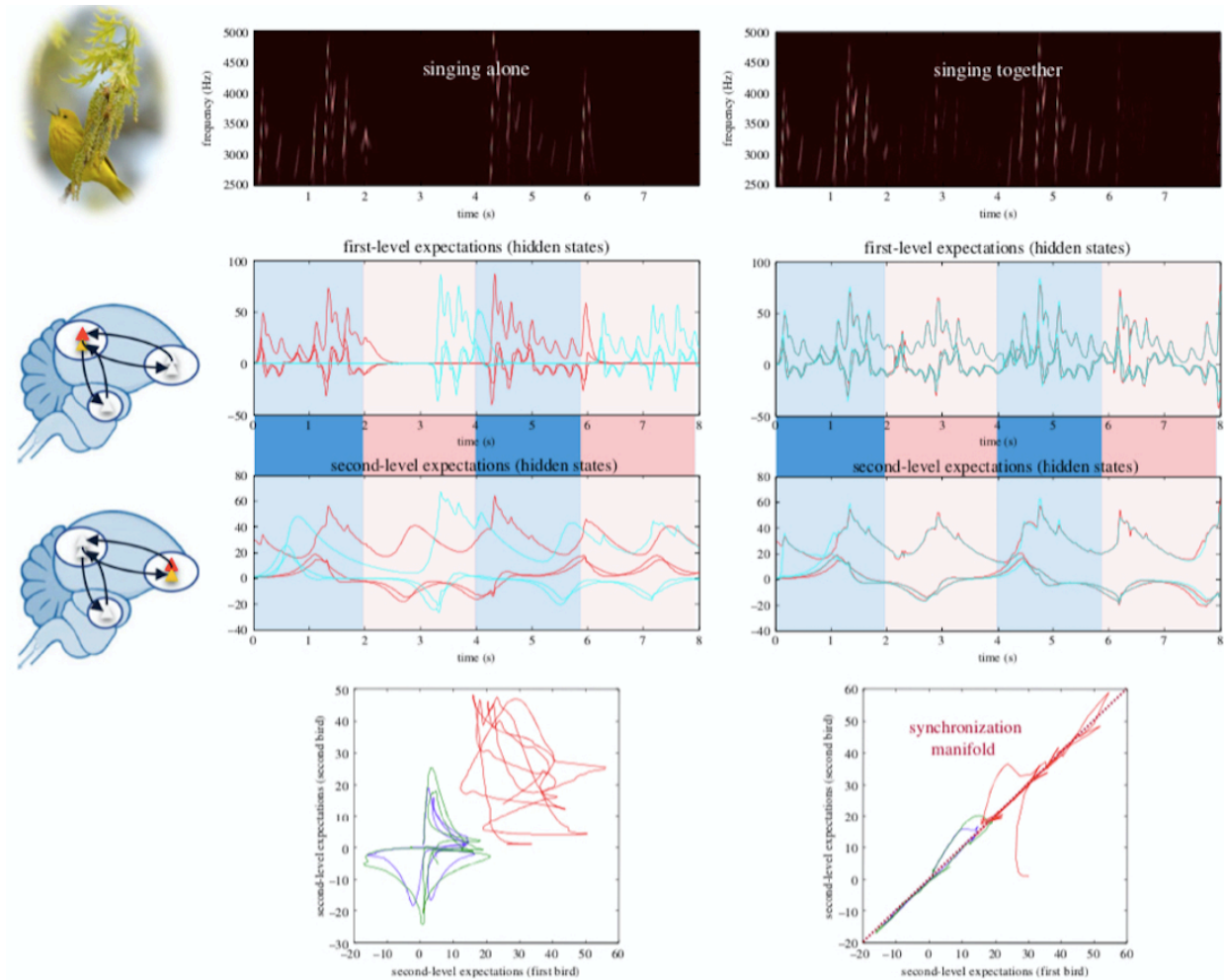

**Fig. 5.** *A simulation of free-energy minimization of the sort implied by the coupled action-perception cycle.* Two birds – endowed with prior expectations about the hidden states generating a shared (birdsong) narrative – sing for 2 s and then listen for a response. The posterior expectations for the first bird are shown in red; and the equivalent expectations for the second bird are shown in blue (both as a function of time). The left panel shows chaotic and uncoupled dynamics when the birds cannot hear each other ('singing alone'), while the right panel shows the synchrony in hidden states that emerges when the birds exchange sensory signals ('singing together'). The different colors correspond to the three hidden states for each bird. When singing alone, the birds cannot hear each other (because they are too far apart). Consequently, the dynamics diverge due to the sensitivity to initial conditions implicit in their (chaotic) generative models. The sonogram heard by the first bird is given in the upper panel. Because this bird can only hear itself, the sonogram reflects the predictions about action based upon its (first- and second-level) posterior expectations. Compare this to the case when the two birds can hear each other ('singing together'). Here, the posterior expectations encoded by internal states show (identical) synchrony at both the sensory and extrasensory levels, as shown in the middle panels (e.g., Pérez, Carreiras, & Duñabeitia, 2017). Note that the sonogram is now continuous over the successive 2 s epochs, because the first bird can hear itself and the second bird. The ensuing synchronization manifold (i.e., the part of the joint state space that contains the generalized synchronization) is shown in the lower panels. These plot the second-level expectations in the second bird against the equivalent expectations in the first. The synchronization manifold for identical synchronization corresponds to the (broken) diagonal line. For details, see Friston & Frith (2015a). Obtained and adapted with permission from Friston & Frith (2015a).

**Figure 6.**

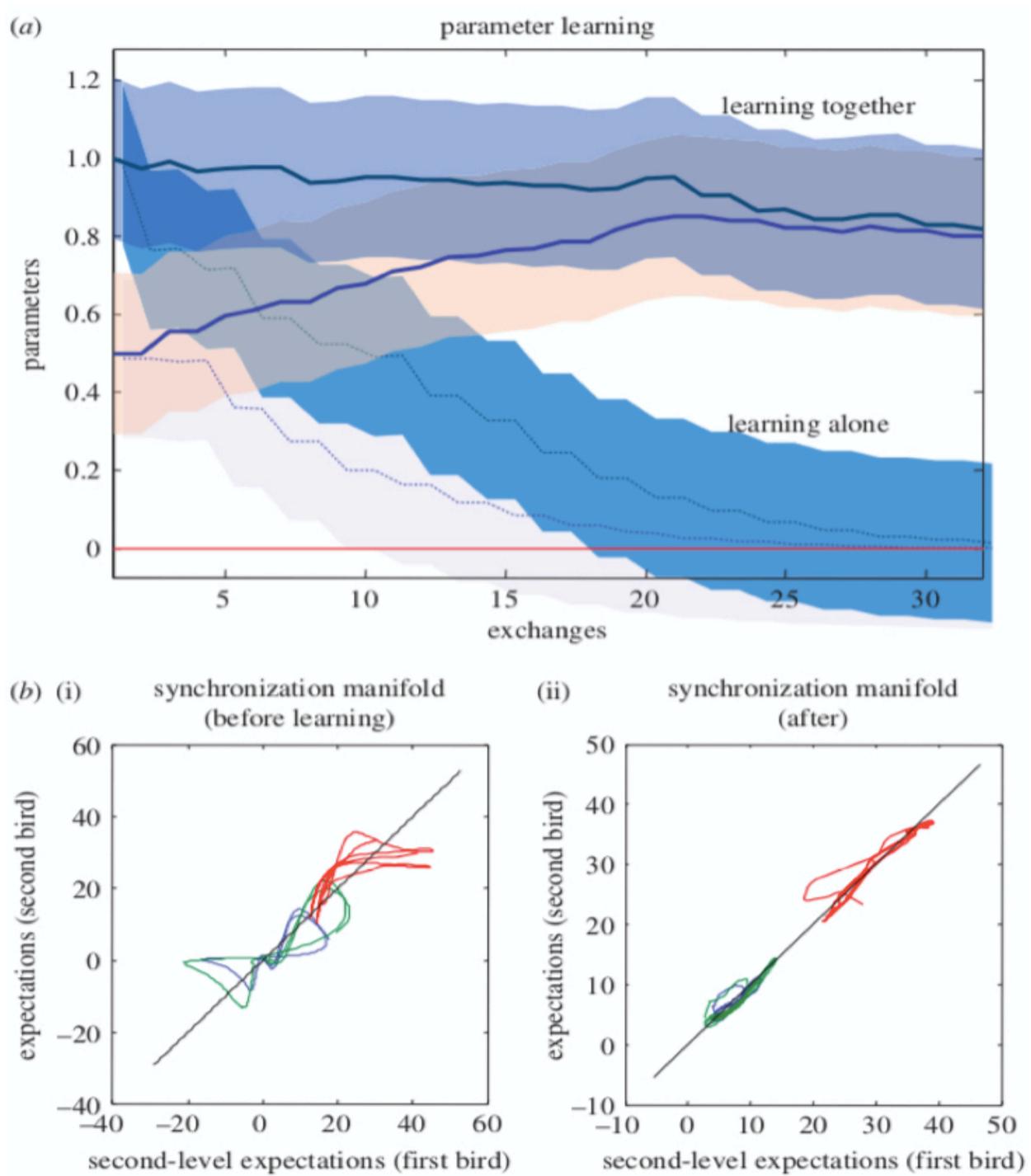

**Fig. 6.** *A duet for one.* This Figure depicts learning and communication via repeated engagement in coupled action-perception cycles in the context of an adaptive prior to align with conspecifics' hidden states. (a) Shows changes in the posterior expectations of an order parameter of the first bird (blue) and second bird (green) determining the chaotic structure of the songs depicted in Figure 5 (by number of reciprocal sensory exchanges). The shaded areas correspond to 90% (prior Bayesian) confidence intervals. The broken lines (and intervals) report the results of the same simulation, but when the birds could not hear each other. (b) Shows the synchronization of posterior expectations encoded by extrasensory areas for the first (i) and subsequent (ii) exchanges, respectively. This synchronization is shown by plotting a mixture of expectations and their temporal derivatives from the second bird against the equivalent expectations of the first bird. This mixture is optimized by assuming a linear mapping between the birds' hidden states. In this example, the second (green) bird had more precise beliefs about its order parameter and, therefore, effectively, 'taught' the first bird. Parameter estimation (learning) converges towards the same value resulting in (generalized) synchrony between the two birds. For details, see Friston & Frith (2015b). Adapted with permission from Friston & Frith (2015b).